\documentclass[aps,prd,a4paper,onecolumn,amsmath,showpacs,superscriptaddress,nofootinbib,preprintnumbers,notitlepage]{revtex4-1}
 
\usepackage{verbatim}
\usepackage[T1]{fontenc}
\usepackage[utf8]{inputenc}
\usepackage[american]{babel}
\usepackage{epsfig}
\usepackage{array}
\usepackage{graphicx,subcaption,caption}
\usepackage{pgfplots}
\usepackage{booktabs}
\usepackage{multirow}
\usepackage{dcolumn}
\usepackage{amsmath}
\usepackage{mathtools}
\usepackage{amsfonts}
\usepackage{amssymb}
\usepackage{epstopdf}
\usepackage{bm}
\usepackage{siunitx}
\usepackage{braket}
\usepackage{enumitem}
\usepackage{soul}
\usepackage{color}
\usepackage{transparent}
\usepackage{pifont}
\usepackage[font={small}]{caption}
\definecolor{navyblue}{rgb}{0.0, 0.0, 0.5}
\definecolor{royalblue}{rgb}{0.25, 0.41, 0.88}
\definecolor{cadmiumgreen}{rgb}{0.0, 0.42, 0.24}
\definecolor{blue-violet}{rgb}{0.54, 0.17, 0.89}
\definecolor{darkviolet}{rgb}{0.58, 0.0, 0.83}
\definecolor{orange(colorwheel)}{rgb}{1.0, 0.5, 0.0}

\usepackage{hyperref}
\hypersetup{
    colorlinks=true, % false: boxed links; true: colored links
    linkcolor=royalblue, % color of internal links (change box color with linkbordercolor)
    citecolor=magenta}

\newcommand\be{\begin{equation}}
\newcommand\ee{\end{equation}}
\newcommand\bea{\begin{eqnarray}}
\newcommand\eea{\end{eqnarray}}

\usepackage{booktabs}
\usepackage{multirow}
\usepackage{dcolumn}
\usepackage{colortbl}

\definecolor{magenta(process)}{rgb}{1.0, 0.0, 0.56}

\definecolor{darkspringgreen}{rgb}{0.09, 0.45, 0.27}

\definecolor{royalblue(web)}{rgb}{0.25, 0.41, 0.88}

\begin{document}

\title{ Constant-roll $f(R)$ inflation compared  with Cosmic Microwave Background anisotropies and swampland criteria}

\author{Mehdi Shokri}
\email{mehdishokriphysics@gmail.com}
\affiliation{Department of Physics, Campus of Bijar, University of Kurdistan, Bijar, Iran}

\author{Mohammad Reza Setare}
\email{rezakord@ipm.ir}
\affiliation{Department of Physics, Campus of Bijar, University of Kurdistan, Bijar, Iran}

\author{Salvatore Capozziello}
\email{capozziello@na.infn.it}
\affiliation{Dipartimento di Fisica E. Pancini", Universit\'a di Napoli Federico II", Via Cinthia, I-80126, Napoli, Italy}
\affiliation{Istituto Nazionale di Fisica Nucleare (INFN), sez. di Napoli, Via Cinthia 9, I-80126 Napoli, Italy}
\affiliation{Scuola Superiore Meridionale, Largo S. Marcellino, I-80138, Napoli, Italy}
\affiliation{Laboratory for Theoretical Cosmology, Tomsk State University of
Control Systems and Radioelectronics (TUSUR), 634050 Tomsk, Russia}

\author{Jafar Sadeghi}
\email{pouriya@ipm.ir}
\affiliation{Department of Physics, University of Mazandaran, P. O. Box 47416-95447, Babolsar, Iran}
\affiliation{School of Physics, Damghan University, P. O. Box 3671641167, Damghan, Iran}

\preprint{}
\begin{abstract}
Inflationary models derived from  $f(R)$ gravity, where the scalaron rolls down with a constant rate from the top  to the minimum  of the effective potential, are considered. Specifically,  we take into account  three  $f(R)$ models \textit{i.e.}\,Starobinsky $R^{2}$, $R^{2p}$ and the logarithmic corrected models. We compare the inflationary parameters derived from the  models with the observational data of CMB anisotropies \textit{i.e.} the Planck and Keck/array datasets in order to find  observational constraints on the parameters space. We find that although our $f(R)$ constant-roll models for $\gamma=0$ show observationally acceptable values of $r$, they do not predict favoured values of the spectral index. In particular, we have $n_{s}>1$ for the Starobinsky $R^{2}$ and   $R^{2p}$ models and $0.996<n_{s}<0.999$ for logarithmic model. Finally, we study the models from the point of view of  Weak Gravity Conjecture adopting  the swampland criteria. 
\\\\
{\bf PACS:} 04.50.$-$h; 98.80.Cq; 98.80.$-$K.
\\{\bf Keywords:} Extended theories of gravity; Constant-roll inflation; Cosmic Microwave Background.
\end{abstract}

\maketitle
\section{Introduction}
Cosmological inflation  is the most straightforward paradigm to describe  the early universe phenomena. In particular,  the inflationary  scalar and tensor perturbations  are unanimously considered  responsible for structure formation and primordial gravitational waves respectively \cite{Guth,Sato:1980yn,Kazanas:1980tx,Linde:1981my,Albrecht:1982wi,Lyth:1998xn}. The simplest inflationary model is based on a single scalar field, the so-called inflaton, rolling down slowly from the top to the minimum point of potential under the slow-roll approximation. Then, the inflationary epoch terminates when inflaton decays to standard particles at the final step of inflation due to the reheating process \cite{Kofman2,Shtanov}. The single field models have been studied broadly in inflationary literature and also have been compared with the observational datasets. Consequently, some of them are restricted or even ruled out \cite{martin} and some models are still viable with the recent high precision observations \cite{staro,barrow,bezrukov,kallosh1}. Moreover, this group of models does not show any non-Gaussianity in their primordial spectrum because of uncorrelated modes of the spectrum \cite{Chen}. In such a case, if the future observations predict non-Gaussianity in the perturbations spectrum, then the single field models will be situated in an unstable status. Recently, a new approach to inflation has been proposed in which inflaton is rolling down with a constant rate \cite{martin2,Motohashi1,Motohashi2}. Hence, $\ddot{\varphi}$ as a non-negligible term can be expressed as
\begin{equation}
\ddot{\varphi}=\beta H\dot{\varphi}
\label{1}
\end{equation}
where $\beta=-(3+\alpha)$ and $\alpha$ is a non-zero parameter. For $\alpha=-3$, the model is reduced to the standard slow-roll. Going beyond of the slow-roll approximation, we can consider   a ultra slow-roll regime where the term of $\ddot{\varphi}$ is non-negligible in the Klein-Gordon equation as $\ddot{\varphi}=3H\dot{\varphi}$. The ultra slow-role models show a limited amount for the non-decaying mode of curvature perturbations \cite{Inoue} and also predict a large $\eta$ but unable to solve the $\eta$ problem introduced in supergravity for the hybrid inflationary models \cite{Kinney}. Moreover, the ultra solutions are located in the non-attractor phase of inflation but reveal a scale-invariance of the scalar perturbation spectrum. Besides, the main problem of ultra models is that the non-Gaussinaity consistency relation of single field models is violated in the presence of ultra conditions through super-Hubble evolution of the scalar perturbation \cite{Namjoo}. The fast-roll models are another class of inflationary models introduced to go   beyond  the slow-roll approximation in which a fast-rolling stage is considered at the start of inflation and can be connected to the standard slow-roll only after a few e-folds \cite{Contaldi,Lello}. 

Constant-roll approach has opened a new window to analyze cosmic inflation due to a constant rate of rolling for inflaton. One can find a wide range of inflationary models investigated in the context of constant-roll idea \cite{Odintsov,Cicciarella,Awad,Anguelova,Ito,Yi,Morse,karam1,Ghersi,Lin,Micu,Oliveros,Motohashi3,Kamali,diego,setare,new2}, in particular,  models coming from  $f(R)$ gravity, where new scalar fields are not required  to drive inflation and one only takes into account  geometric extensions of the Hilbert-Einstein action. In \cite{Nojiri}, the authors  considered two different approaches. First, they  studied the constant-roll  evolution of a scalar-tensor theory in the presence of $f(R)$ gravity and, as  second approach, they  applied the constant-roll condition directly into $f(R)$ gravity. Generalizations of the costant-roll condition in scalar-tensor gravity \cite{Oikonomou:2021yks}, in Gauss-Bonnet gravity \cite{Oikonomou:2020oil} and in connection to reheating in $f(R)$ gravity \cite{Oikonomou:2017bjx} have been also developed. On the other hand, 
logarithmic-corrected $R^2$ gravity  in  presence of
Kalb-Ramond fields has been considered in \cite{Elizalde:2018now}. This model, i.e logarithmic
$f(R)$ + Kalb-Ramond, seems to be consistent with the observable values of
inflationary parameters. Specifically, the authors consider  both $\gamma = 0$  and non-zero $\gamma$. Finally, the constant slow-roll condition in Palatini formalism has been  taken into account in
\cite{Antoniadis:2020dfq}.
In \cite{Motohashi9}, the authors  introduced the constant-roll condition in the Jordan frame, and by calculating the potential of scalaron in the Einstein frame, they  obtained the form of $f(R)$ in the Jordan frame.

The main aim of the present manuscript is to find the observational constraints from Cosmic Microwave Background (CMB) anisotropies  on the parameters space of some  $f(R)$ models introduced in the context of the constant-roll inflation. In practice, we use as CMB data, the Planck and the BICEP2/Keck array data releases \cite{cmb, bicep}. Also, we compare the predictions of the models with observations in order to find the effects of the new approach. As a secondary purpose, we investigate the $f(R)$ constant-roll inflationary models in view of swampland conditions since $f(R)$ models contains some stringy-like corrections to GR and also $f(R)$ gravity can be realized as a natural class of theories emerging from the string landscape because of the existence of Noether symmetries in this class of extended theories of gravity \cite{Gionti, Benetti}. 

First, we study the Starobinsky $R^{2}$ model as the most well-known inflationary model in $f(R)$ gravity  containing a quadratic term coming from higher-order curvature  added to Ricci scalar in the Hilbert-Einstein  action as ${\displaystyle f(R)=R+\frac{R^{2}}{6M^{2}}}$ \cite{staro}. The Starobinsky model is also well-known as one of the most successful inflationary models  being in good agreement with current observations. Hence, it is now considered as a "target" model for several future CMB experiments as, for example, the Simons Observatory \cite{Aguirre}, CMB-S4 \cite{Abazajian}, and the LiteBIRD satellite experiment \cite{Suzuki}. Also, $R^{2}$ model  has  been proposed as one of the possible alternatives to the cosmological constant of the concordance $\Lambda$CDM model \cite{a,b,c,c,d,d,e,f,g,h,i,k}. Despite the mentioned achievements, the Starobinsky model predicts a very tiny tensor-to scalar ratio $r\simeq0.003$ for 60 e-folds which is out of the current and probably  future observations. The goal of these future experiments is therefore to improve the  experimental sensitivity to measure such a signal with enough statistical significance with $\delta s<0.001$. Since the predicted value of $r$ is the first approximation, we might consider some theoretical corrections to the Starobinsky model in order to improve the obtained value. Therefore, we focus on a generalized form of the Starobinsky inflation, the so-called $R^{2p}$ model (with $p\approx1$). These inflationary models were first proposed by \cite{Schmidt,Maeda} in the context of higher-derivative theories and subsequently were applied to inflation providing a simple and elegant generalization of the $R^{2}$ inflation \cite{Muller,Felice,Ringeval,Trotta,new0}. Moreover, we consider the logarithmic corrected Starobinsky model $f(R)= R+\lambda R^{2}+\upsilon R^{2}\ln R$ where logarithmic corrections come from  quantum gravity.  Logarithmic $f(R)$ gravity can be considered a prototype model  with quantum corrections, able to describe  primordial and current accelerated expansions of the universe. Inflationary examples of logarithmic $f(R)$ models can be found in Refs. \cite{Starobinsky3,Sotiriou,Felice,a1,a2,a3,a4,a5,a6,a7,a8,a9,a11,a12,a15,a16,a17,a18}. For other cosmological situations see \cite{a21,a22,a23,a25,a26,a27,a29,a30,a32,a34,a35}.

The above discussion motivates us to arrange the paper as follows. In \S II, we introduce $f(R)$ gravity and its main properties in both Jordan and Einstein frames with the connecting relations of parameters in the two frames.  \S III is devoted to the  study of the constant-roll inflation in $f(R)$ gravity. In \S IV, we focus on the $f(R)$ models and obtain the corresponding potentials and also the inflationary parameters. In \S V, we analyze the obtained results  by comparing with the observational datasets coming from the Planck and the BICEP2/Keck array satellites. Also, we investigate the models from the viewpoint of the Weak Gravity Conjecture using the swampland criteria. In \S VI, we conclude the analysis of the models and draw our future outlooks.

\section{$f(R)$ Gravity in Jordan and Eistein frame}
Let us start with the general  form of  $f(R)$ gravity action
\begin{equation}
S=\int{d^{4}x\sqrt{-g}\frac{f(R)}{2}}+\int{d^{4}x\mathcal{L}_{M}(g_{\mu\nu},\Psi_{M})}
\label{2}
\end{equation}
where $g$ is the determinant of the metric $g_{\mu\nu}$, $R=g^{\mu\nu}R_{\mu\nu}$ is the Ricci scalar as gravitational sector and $\mathcal{L}_{M}$ is the Lagrangian of matter fields $\Psi_{M}$ filling the universe as perfect fluid \cite{Report,Mantica,CANTATA}. Also, here we suppose $\kappa^{2}\equiv8\pi G=1$. By varying the action (\ref{2}) respect to the metric, the Einstein field equation is given by 
\begin{equation}
\mathcal{G}_{\mu\nu}\equiv FR_{\mu\nu}-\frac{g_{\mu\nu}f}{2}-\nabla_{\mu}\nabla_{\nu}F+g_{\mu\nu}\Box F=T_{\mu\nu}^{m}
\label{3}
\end{equation} 
where $F$, d'Alembert $\Box$ and the energy-momentum tensor of matter fields $T_{\mu\nu}^{m}$ are expressed by \begin{equation}
F\equiv\frac{df(R)}{dR},\hspace{1cm}\Box=\frac{1}{\sqrt{-g}}\partial_{\nu}[\sqrt{-g}g^{\mu\nu}\partial_{\mu}],\hspace{1cm} T_{\mu\nu}^{m}=-\frac{2}{\sqrt{-g}}\frac{\delta\mathcal{L}_{M}}{\delta g_{\mu\nu}}.
\label{4}
\end{equation} 
The conservation law of the energy-momentum tensor $\nabla^{\mu} T_{\mu\nu}^{m}=0$ is valid when $\nabla^{\mu}\mathcal{G}_{\mu\nu}=0$. Also, the field equation (\ref{3}) can be rewritten in terms of the Einstein tensor $G_{\mu\nu}$ as
\begin{equation}
G_{\mu\nu}\equiv R_{\mu\nu}-\frac{1}{2}g_{\mu\nu}R=T_{\mu\nu}^{m}+T_{\mu\nu}^{c}
\label{5}
\end{equation}
where $T_{\mu\nu}^{c}$ including the terms coming from $f(R)$ modifications takes the following form 
\begin{equation}
T_{\mu\nu}^{c}\equiv (1-F)R_{\mu\nu}+\frac{1}{2}(f-R)g_{\mu\nu}+\nabla_{\mu}\nabla_{\nu}F-g_{\mu\nu}\Box F
\label{6}
\end{equation}
and the conservation law of the energy-momentum tensor $\nabla^{\mu} T_{\mu\nu}^{m}=0$ is valid as a consequence of the Bianchi identities  $\nabla^{\mu} G_{\mu\nu}=0$ if $\nabla^{\mu} T_{\mu\nu}^{c}=0$. Obviously, in the case of $f(R)=R$, both approaches reduce to Einstein gravity. In the following we consider a spatially flat universe described by a Friedmann-Robertson-Walker (FRW) metric as 
\begin{equation}
ds^{2}=-dt^{2}+a(t)^{2}(dx^{2}+dy^{2}+dz^{2})   
\label{7}
\end{equation}
where $t$ and $a$ depict to cosmic time and scale factor, respectively. By using the above metric and the definition of the energy-momentum tensor of perfect fluid as gravitational and matter sectors of the Eq. (\ref{3}), we obtain the dynamical equations as
\begin{equation}
3FH^{2} = \frac{(F R-f)}{2}-3H\dot{F}+\rho_{m},\hspace{1cm} 2F\dot{H}=-\ddot{F}+H\dot{F}-(\rho_{m} +P_{m}) 
\label{8}
\end{equation}
\noindent where $H\equiv\frac{\dot{a}}{a}$ is Hubble parameter and dot represents time derivation. Moreover, $\rho_{m}$ and $P_{m}$ are energy density and pressure of the perfect fluid. In the rest of the paper, we confine ourselves to inflationary analysis in $f(R)$ gravity by removing the role of matter fields in the main action (\ref{2}) since we don't require to consider any type of matter in order to push inflation. Hence, the inflationary action in $f(R)$ gravity takes the simple form 
\begin{equation}
S=\int{d^{4}x\sqrt{-g}\frac{f(R)}{2}}
\label{9}
\end{equation}
with the dynamical equations
\begin{equation}
3FH^{2} = \frac{(F R-f)}{2}-3H\dot{F},\hspace{1cm} 2F\dot{H}=-\ddot{F}+H\dot{F}.
\label{10}
\end{equation}
By using the conformal transformation as a useful mathematical tool, we can move from Jordan frame (main frame) to Einstein frame in order to escape from difficulties of $f(R)$ gravity. The metrics in the two frames are connected with  
\begin{equation}
\hat{g}_{\mu\nu}=\Omega^{2}g_{\mu\nu}\quad\quad with \quad\quad \Omega^{2}=F=e^{\sqrt{\frac{2}{3}}\varphi}. 
\label{11}
\end{equation}
Hence, the form of action in the Einstein frame takes the standard form as 
\begin{equation}
S_{E}=\int{d^{4}x\bigg(\hat{R}-\frac{1}{2}g^{\mu\nu}\partial_{\mu}\varphi\partial_{\nu}\varphi-V(\varphi)}\bigg)
\label{12}
\end{equation}
where hat denotes to the parameters in the Einstein frame. Clearly, in the Einstein frame, we deal with a scalar field $\varphi$ so that called scaleron with the potential
\begin{equation}
V(\varphi)=\frac{RF-f}{2F^{2}}.
\label{13}
\end{equation}
Conversely, the Ricci scalar and the function of $f(R)$ in the Jordan frame can be connected to the potential in the Einstein frame by
\begin{equation}
R=e^{\sqrt{\frac{2}{3}}\varphi}\bigg(\sqrt{6}\frac{\partial V}{\partial\varphi}+4V\bigg),\quad\quad\quad f(R)=e^{2\sqrt{\frac{2}{3}}\varphi}\bigg(\sqrt{6}\frac{\partial V}{\partial\varphi}+2V\bigg).
\label{14}
\end{equation}
Moreover, the dynamical equations in the Einstein frame are defined by 
\begin{equation}
\hat{H}^{2}=\frac{1}{3}\bigg(\frac{1}{2}(\frac{d\varphi}{d\hat{t}})^{2}+V(\varphi)\bigg),\hspace{1cm} \frac{d\hat{H}}{d\hat{t}}=-\frac{1}{2}(\frac{d\varphi}{d\hat{t}})^{2},\hspace{1cm} \frac{d^{2}\varphi}{d\hat{t}^{2}}+3\hat{H}(\frac{d\varphi}{d\hat{t}})
+\frac{\partial V}{\partial\varphi}=0.
\label{15}
\end{equation}
Also, the parameters in the two frames are connected by 
\begin{equation}
H=e^{\frac{\varphi}{\sqrt{6}}}\bigg(\hat{H}-\frac{1}{\sqrt{6}}\frac{d\varphi}{d\hat{t}}\bigg),\hspace{1cm} \frac{d\varphi}{dt}=e^{\frac{\varphi}{\sqrt{6}}}\frac{d\varphi}{d\hat{t}},\hspace{1cm} \frac{d^{2}\varphi}{dt^{2}}=e^{\sqrt{\frac{2}{3}}\varphi}\bigg(\frac{d^{2}\varphi}{d\hat{t}^{2}}+\frac{1}{\sqrt{6}}(\frac{d\varphi}{d\hat{t}})^{2}\bigg).
\label{16}
\end{equation}
Note that the above expressions are obtained due to $dt=e^{-\frac{\varphi}{\sqrt{6}}}d\hat{t}$ and $a=e^{-\frac{\varphi}{\sqrt{6}}}\hat{a}$.

\section{Constant-roll $f(R)$ inflation}
Deviations from slow-roll approximation can be found when we restrict our attention to ultra slow-roll inflationary models in which $\alpha=0$ or even fast-roll models in which a fast-rolling period is considered at the beginning of inflation and then will be transited to the slow-roll stage just after a few e-folds. As a new and more complete approach, we can introduce constant-roll inflation in which inflaton rolls down with a constant rate from top of the potential to the minimum point. This viewpoint has been proposed to provide non-Gaussianity properties in the single field models since these models do not predict any non-Gaussianity in the context of slow-roll approximation. Recently, diverse inflationary models have been investigated in the presence of the constant-roll condition. In this paper, we focus on some inflationary models motivated by $f(R)$ gravity. Hence, we consider a natural generalization of the constant-roll condition (\ref{1}) in $f(R)$ gravity expressed in the Jordan frame as
\begin{equation}
\ddot{F}=\beta H\dot{F}
\label{17}
\end{equation}
which is reduced to the standard slow-roll approximation for $\beta=0$. Since the inflationary potential of $f(R)$ is defined in the Einstein frame, we follow the mechanism in the Einstein frame using the connecting relations between two frames (\ref{16}). By the definition of $F$ (\ref{11}), the constant-roll condition (\ref{17}) in the Einstein frame is rewritten by
\begin{equation}
\frac{d^{2}\varphi}{d\hat{t}^{2}}+\frac{3+\beta}{\sqrt{6}}\bigg(\frac{d\varphi}{d\hat{t}}\bigg)^{2}-\beta\hat{H}\frac{d\varphi}{d\hat{t}}=0.
\label{18}
\end{equation}
Then, by using the Einstein equation (\ref{15}), the above equation takes the following form
\begin{equation}
\frac{d\hat{H}}{d\varphi}\bigg(\frac{d^{2}\hat{H}}{d\varphi^{2}}+\frac{3+\beta}{\sqrt{6}}\frac{d\hat{H}}{d\varphi}+\frac{\beta}{2}\hat{H}\bigg)=0
\label{19}
\end{equation}
which includes two separated cases. The first case $\frac{d\hat{H}}{d\varphi}=0$ is associated to the case of $\hat{H} = const.$ which leads to $V= const.$ and then $f(R) = R - const.$ The second case is a second order differential equation with a general solution
\begin{equation}
\hat{H}(\varphi)=C\bigg(\gamma F^{-\frac{3}{2}}+F^{-\frac{\beta}{2}}\bigg)   
\label{20}
\end{equation}
and by using the Friedmann equation (\ref{15}), the potential and the evolution of the scaleron are obtained as
\begin{equation}
V(\varphi)=\frac{3-\beta}{3}C^{2}\bigg(6\gamma F^{-\frac{(3+\beta)}{2}}+(3+\beta)F^{-\beta}\bigg),\quad\quad\quad\frac{d\varphi}{d\hat{t}}=-2\frac{d\hat{H}}{d\varphi}=\frac{2}{\sqrt{6}}\bigg(3\gamma F^{\frac{-3}{2}}+\beta F^{-\frac{\beta}{2}}\bigg)
\label{21}
\end{equation}
where $F=e^{\sqrt{\frac{2}{3}}\varphi}$. The parameters $C$ and $\gamma$ are the integration constants mass dimension 1 and dimensionless, respectively. The amplitude of $C$ can be fixed by the CMB normalization and we follow the paper in the unit where $C=1$. Also, the value of $\gamma$ can be normalized by redefining $C$ and $\varphi$ so that it can be considered for three value $+1, 0, -1$. In ref. \cite{Motohashi9}, the authors claimed that only the parameters region with $\beta=-0.02$ and $\gamma=-1$ are viable, cosmologically. In other words, these parameters are situated in a region where inflaton shows an attractor-like behavior. By combination of the potential (\ref{21}) and the connection relations (\ref{14}), the Ricci scalar takes the following form
\begin{equation}
R=(\beta-3)\bigg(2\gamma(\beta-1)e^{-\frac{(1+\beta)\varphi}{\sqrt{6}}}+\frac{2}{3}(\beta-2)(\beta+3)e^{\frac{2(1-\beta)\varphi}{\sqrt{6}}}\bigg).
\label{22}
\end{equation}
Now, we can calculate the slow-roll parameters as
\begin{equation}
\epsilon\equiv\frac{1}{2}\bigg(\frac{V'}{V}\bigg)^{2}=\frac{(3+\beta)^{2}F'^{2}\bigg(3\gamma F^{\frac{-(5
+\beta)}{2}}+\beta F^{-(\beta+1)}\bigg)^{2}}{2\bigg(6\gamma F^{\frac{-(3+\beta)}{2}}+(3+\beta)F^{-\beta}\bigg)^{2}},
\label{23}
\end{equation}
\begin{equation}
\eta\equiv\frac{V''}{V}=\frac{-(3+\beta)\bigg(F''(3\gamma F^{\frac{-(5+\beta)}{2}}+\beta F^{-(\beta+1)})-F'^{2}(\frac{3\gamma(5+\beta)}{2}F^{-\frac{(7+\beta)}{2}}+\beta(1+\beta)F^{-(2+\beta)})\bigg)}{\bigg(6\gamma F^{\frac{-(3+\beta)}{2}}+(3+\beta)F^{-\beta}\bigg)},
\label{24}
\end{equation}
\begin{eqnarray}
&\!&\!\zeta^{2}\equiv\frac{V'V'''}{V^{2}}=\frac{(3+\beta)^{2}F'\bigg(3\gamma F^{\frac{-(5
+\beta)}{2}}+\beta F^{-(\beta+1)}\bigg)}{\bigg(6\gamma F^{\frac{-(3+\beta)}{2}}+(3+\beta)F^{-\beta}\bigg)^{2}}\times\nonumber\\&\!&\!
\times\bigg(F'''(3\gamma F^{\frac{-(5+\beta)}{2}}+\beta F^{-(\beta+1)})-3F''F'(\frac{3\gamma(5+\beta)}{2}F^{-\frac{(7+\beta)}{2}}+\nonumber\\&\!&\!
+\beta(1+\beta)F^{-(2+\beta)})+F'^{3}(\frac{3\gamma(5+\beta)(7+\beta)}{4}F^{-\frac{(9+\beta)}{2}}+\beta(1+\beta)(2+\beta)F^{-(3+\beta)})\bigg)
\label{25}
\end{eqnarray}
where prime implies to the derivation with respect to the scalar field $\varphi$ and inflation ends when the condition $\epsilon = 1$ or $\eta = 1$ is fulfilled. Also, we can apply the swampland conditions $\sqrt{2\epsilon}\geq c$ and $|\eta|\leq-c'$,  first derived in \cite{Garg:2018reu} and then in  \cite{K1},  for the $f(R)$ models by
\begin{equation}
\frac{(3+\beta)F'\bigg(3\gamma F^{\frac{-(5
+\beta)}{2}}+\beta F^{-(\beta+1)}\bigg)}{\bigg(6\gamma F^{\frac{-(3+\beta)}{2}}+(3+\beta)F^{-\beta}\bigg)}\geq c,
\label{26}
\end{equation}
\begin{equation}
\Bigg|\frac{-(3+\beta)\bigg(F''(3\gamma F^{\frac{-(5+\beta)}{2}}+\beta F^{-(\beta+1)})-F'^{2}(\frac{3\gamma(5+\beta)}{2}F^{-\frac{(7+\beta)}{2}}+\beta(1+\beta)F^{-(2+\beta)})\bigg)}{\bigg(6\gamma F^{\frac{-(3+\beta)}{2}}+(3+\beta)F^{-\beta}\bigg)}\Bigg|\leq c'
\label{27}
\end{equation}
where $c$ and $c'$ are unit orders. Based on our purpose, we need to calculate the number of e-folds in the Einstein frame using 
\begin{equation}
N\equiv\int^{\varphi_{i}}_{\varphi_{f}}{\frac{1}{\sqrt{2\epsilon}}}d\varphi=\int^{\varphi_{i}}_{\varphi_{f}}\frac{\bigg(6\gamma F^{\frac{-(3+\beta)}{2}}+(3+\beta)F^{-\beta}\bigg)}{(3+\beta)F'\bigg(3\gamma F^{\frac{-(5+\beta)}{2}}+\beta F^{-(\beta+1)}\bigg)}
d\varphi
\label{28}
\end{equation}
where the subscribes "$i$" and "$f$" denote the value of the scaleron field at the beginning and the end of inflation, respectively. The spectral parameters \textit{i.e.} the first order of spectral index, the first order of running spectral index and the tensor-to-scalar ratio can be obtained by  
\begin{equation}
n_{s}=1-6\epsilon+2\eta,\quad\quad\quad
\alpha_{s}=\frac{dn_{s}}{d\ln k}=16\epsilon\eta-24\epsilon^{2}-2\zeta^{2},\quad\quad\quad r=16\epsilon.
\label{29}  
\end{equation}
\section{$f(R)$ models}
In this section, we shall examine the constant-roll inflation for some  $f(R)$ models \textit{i.e.} Starobinsky $R^{2}$ , $R^{2p}$ and logarithmic corrected models. 
\begin{figure*}[!hbtp]
	\centering
	\includegraphics[width=.28\textwidth,keepaspectratio]{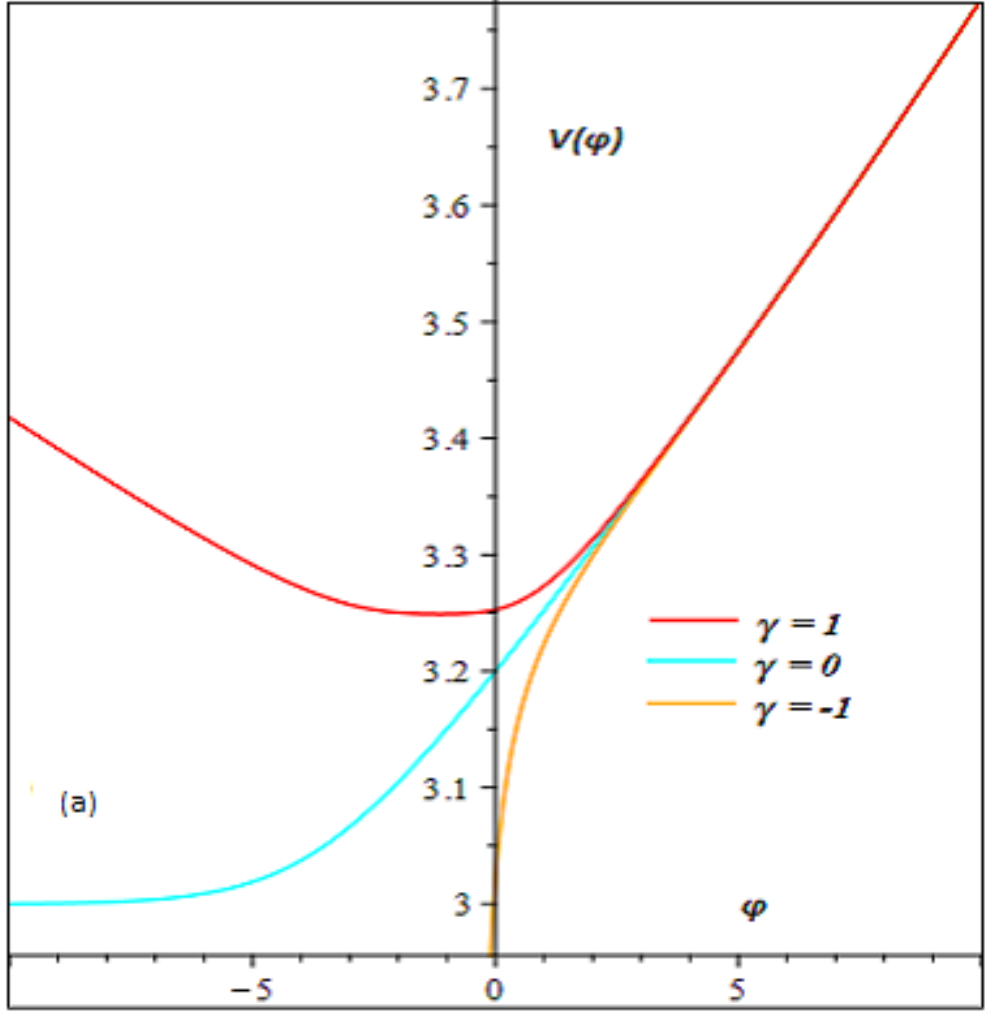}
	\hspace{0.5cm}
	\includegraphics[width=.28\textwidth,keepaspectratio]{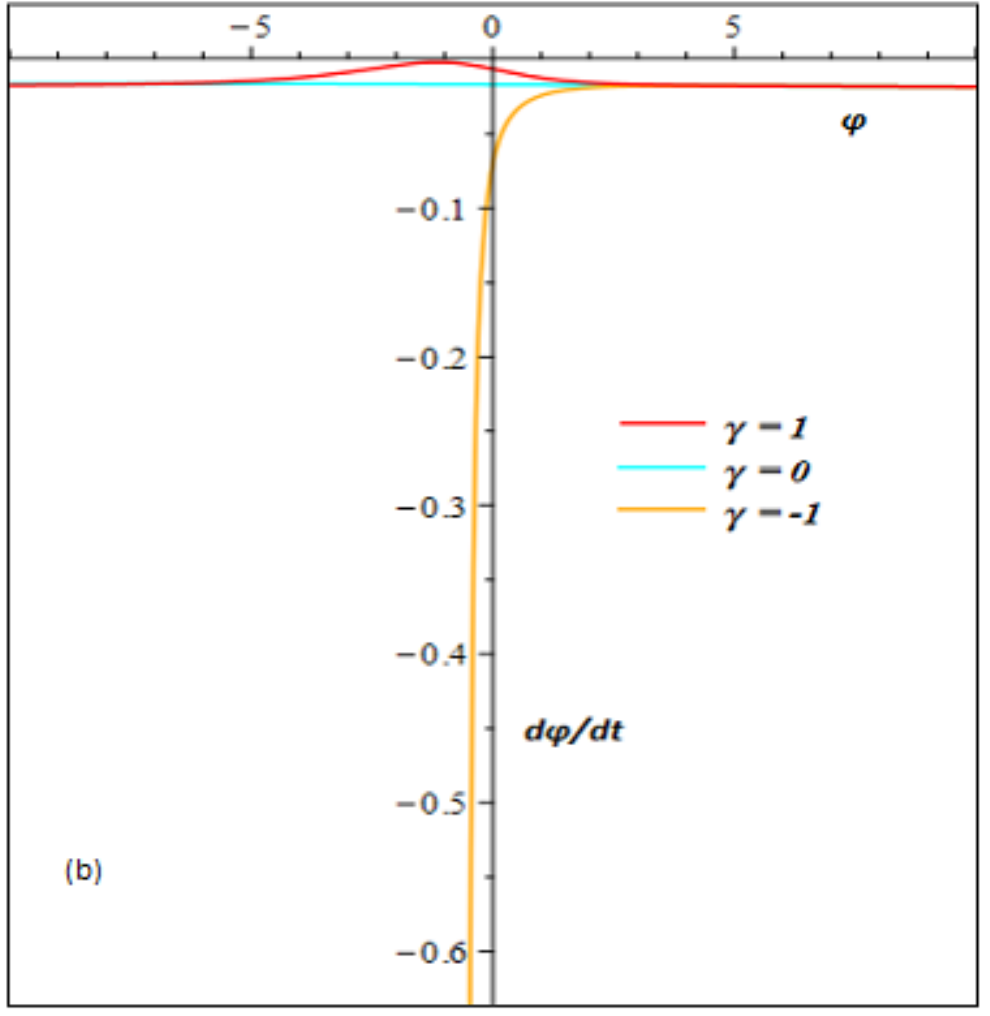}
	\hspace{0.5cm}
	\includegraphics[width=.28\textwidth,keepaspectratio]{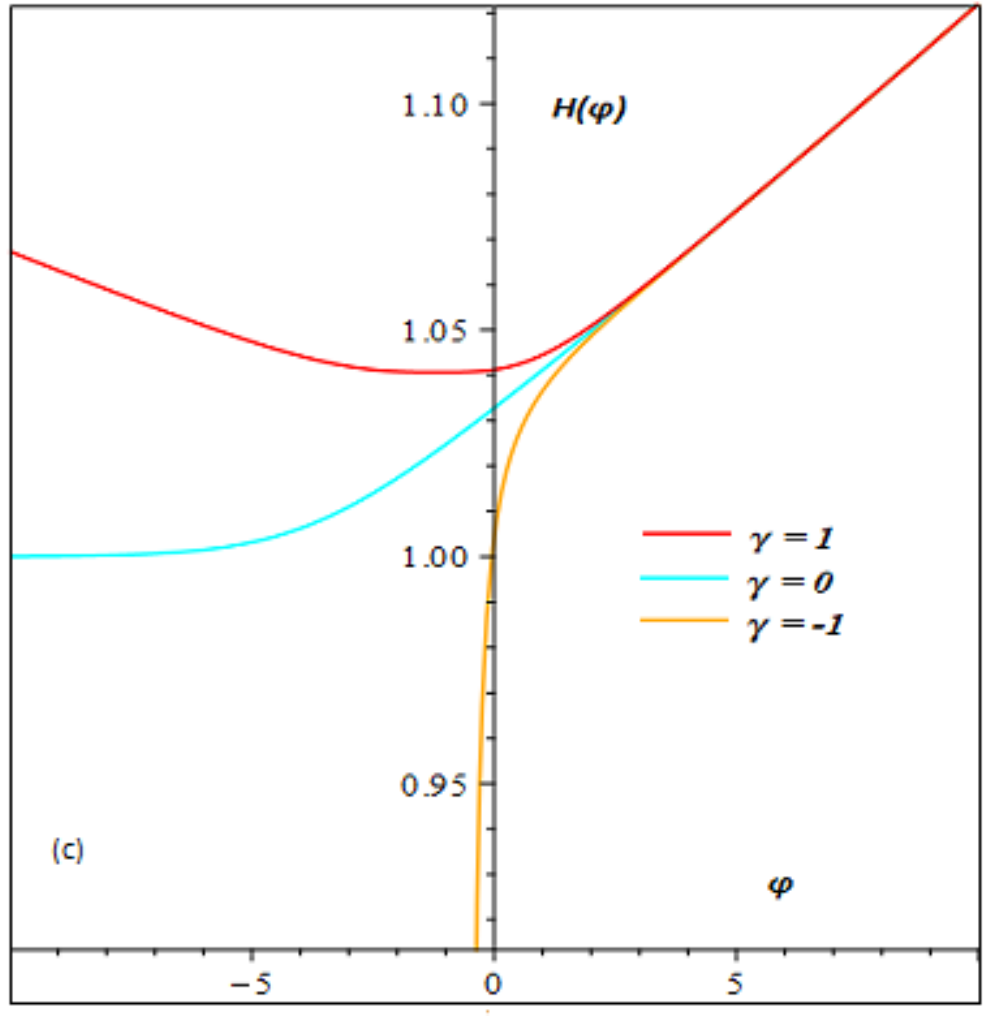}
	\caption{The potential, evolution of scaleron and Hubble parameter  (\ref{32}) plotted versus $\varphi$ for $\gamma=+1, 0, -1$ with $\beta=-0.02$ and $\lambda=1$.}
	\label{fig1}
\end{figure*} 
\subsection{Starobinsky $R^{2}$ model}
First, we start with the Starobinsky $R^{2}$ model as the most successful inflationary model which contains a quadratic term coming from higher-order curvature given by
\begin{equation}
f(R)=R+\lambda R^{2} 
\label{30}
\end{equation}
where $\lambda$ is an arbitrary constant. By using the form of the Ricci scalar (\ref{22}),  $F=\frac{df}{dR}$ for the Starobinsky model can be driven by
\begin{equation}
F_{1}=1+2\lambda(\beta-3)\bigg(2\gamma(\beta-1)e^{-\frac{(1+\beta)\varphi}{\sqrt{6}}}+\frac{2}{3}(\beta-2)(\beta+3)e^{\frac{2(1-\beta)\varphi}{\sqrt{6}}}\bigg).
\label{31}
\end{equation}
Now the Hubble parameter (\ref{20}), the potential of scalar field and the evolution of the scalar field (\ref{21}) are given by
\begin{equation}
\hat{H}_{1}(\varphi)=\bigg(\gamma F_{1}^{-\frac{3}{2}}+F_{1}^{-\frac{\beta}{2}}\bigg),\quad V_{1}(\varphi)=\frac{3-\beta}{3}\bigg(6\gamma F_{1}^{-\frac{(3+\beta)}{2}}+(3+\beta)F_{1}^{-\beta}\bigg),\quad\frac{d\varphi_{1}}{d\hat{t}}=\frac{2}{\sqrt{6}}\bigg(3\gamma F_{1}^{\frac{-3}{2}}+\beta F_{1}^{-\frac{\beta}{2}}\bigg).
\label{32}
\end{equation}
Now let's review the appropriate plots of the Starobinsky $R^{2}$ model. The panel (a) of Fig. \ref{1} presents the behaviour of scaleron potential (\ref{32}) for three values of $\gamma = +1, 0, -1$ with $\beta=-0.02$ and $\lambda=1$. For $\gamma=+1$ the potential rolls down to a minimum point and then tilts upwards. For $\gamma=0$ the potential moves in a steady manner from negative points of scaleron and then gradually shows an increasing behavior. For $\gamma=-1$, the potential starts moving from the origin and tilts upward with a decreasing slope. Besides this information, the panel (a) also reveals that all three values of $\gamma$ act the same with a constant rate at the last steps of inflation. The panels (b) and (c) of Fig. \ref{fig1} show the phase diagram of scaleron and the behaviour of the Hubble parameter (\ref{32}) for the values $\gamma= +1, 0, -1$  with $\beta=-0.02$ and $\lambda=1$, respectively. Notice that in the rest of paper, we study the inflationary paradigm in the case of $\gamma=0$. Now let's find the slow-roll parameters (\ref{23} - \ref{25}) of the Starobinsky $R^{2}$ model for the case of $\gamma=0$ by 
\begin{equation}
\epsilon=\frac{\beta^{2}}{2}\bigg(\frac{F'_{1}}{F_{1}}\bigg)^{2},\quad\eta=-\beta\bigg(\frac{F_{1}''}{F_{1}}-(\beta+1)(\frac{F_{1}'}{F_{1}})^{2}\bigg),\quad\zeta^{2}=\beta^{2}\frac{F_{1}'}{F_{1}^{2}}\bigg(F_{1}'''-3(\beta+1)\frac{F_{1}''F_{1}'}{F_{1}}+(\beta+1)(\beta+2)\frac{F_{1}'^{3}}{F_{1}^{2}}\bigg).
\label{33}    
\end{equation}
Also, the number of e-folds (\ref{28}) of the model can be found as 
\begin{equation}
N=\int^{\varphi_{i}}_{\varphi_{f}}\frac{1}{\beta}\frac{F_{1}}{F'_{1}}d\varphi=\frac{9}{8\lambda\beta(\beta-1)^{2}(\beta-2)(3-\beta)(\beta+3)}\bigg(e^{\frac{2(\beta-1)\varphi_{i}}{\sqrt{6}}}-e^{\frac{2(\beta-1)\varphi_{f}}{\sqrt{6}}}\bigg)+\frac{\sqrt{6}}{2\beta(1-\beta)}(\varphi_{i}-\varphi_{f})
\label{34}    
\end{equation}
and then by neglecting the value of scalar field at the end of inflation ($\varphi_{f}\ll\varphi_{i}$), the above expression is reduced to
\begin{equation}
N\simeq\frac{9}{8\lambda\beta(\beta-1)^{2}(\beta-2)(3-\beta)(\beta+3)}e^{\frac{2(\beta-1)\varphi_{i}}{\sqrt{6}}}.
\label{35}
\end{equation}
Now using the slow-roll parameters (\ref{33}) and the above expression for the number of e-folds, the inflationary parameters (\ref{29}) of the Starobinsky $R^{2}$ model can be calculated as shown in the appendix (\ref{a1} - \ref{a3}).
\begin{figure*}[!hbtp]
	\centering
	\includegraphics[width=.28\textwidth,keepaspectratio]{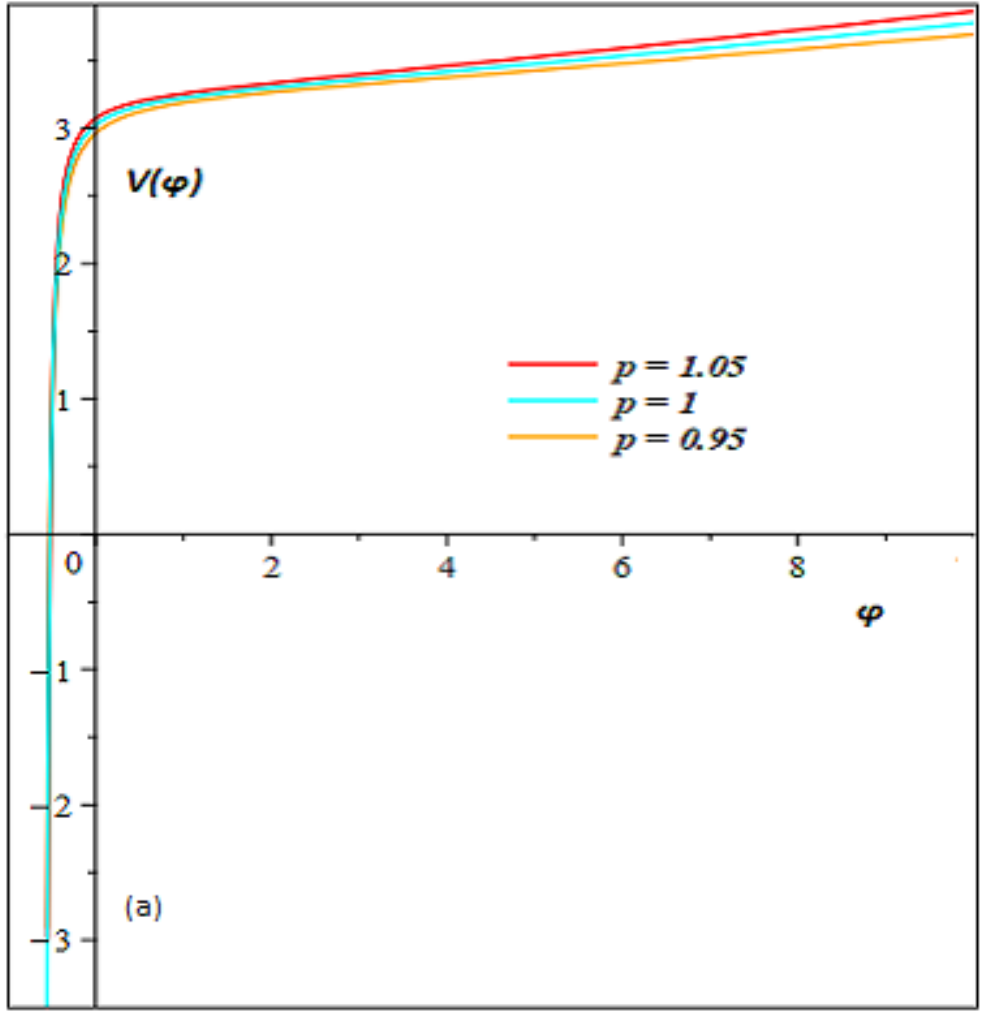}
	\hspace{0.5cm}
	\includegraphics[width=.28\textwidth,keepaspectratio]{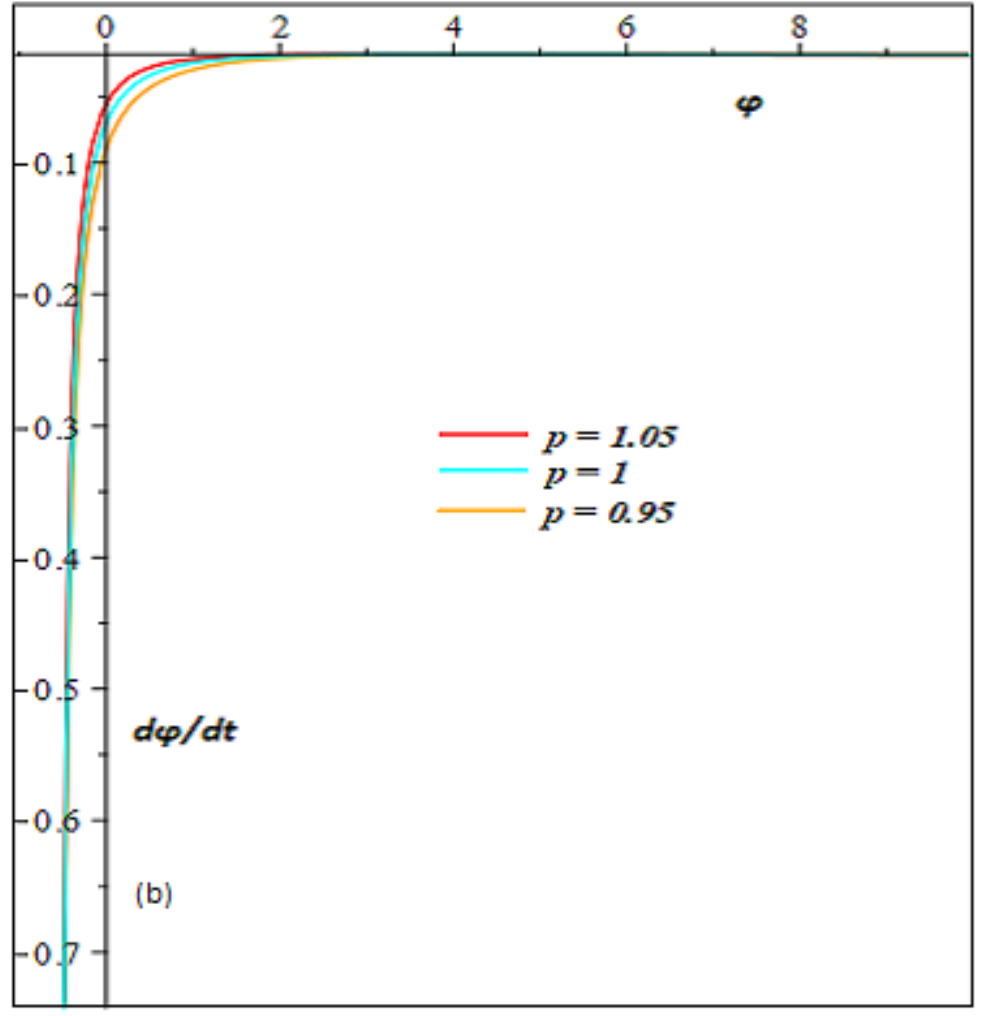}
	\hspace{0.5cm}
	\includegraphics[width=.28\textwidth,keepaspectratio]{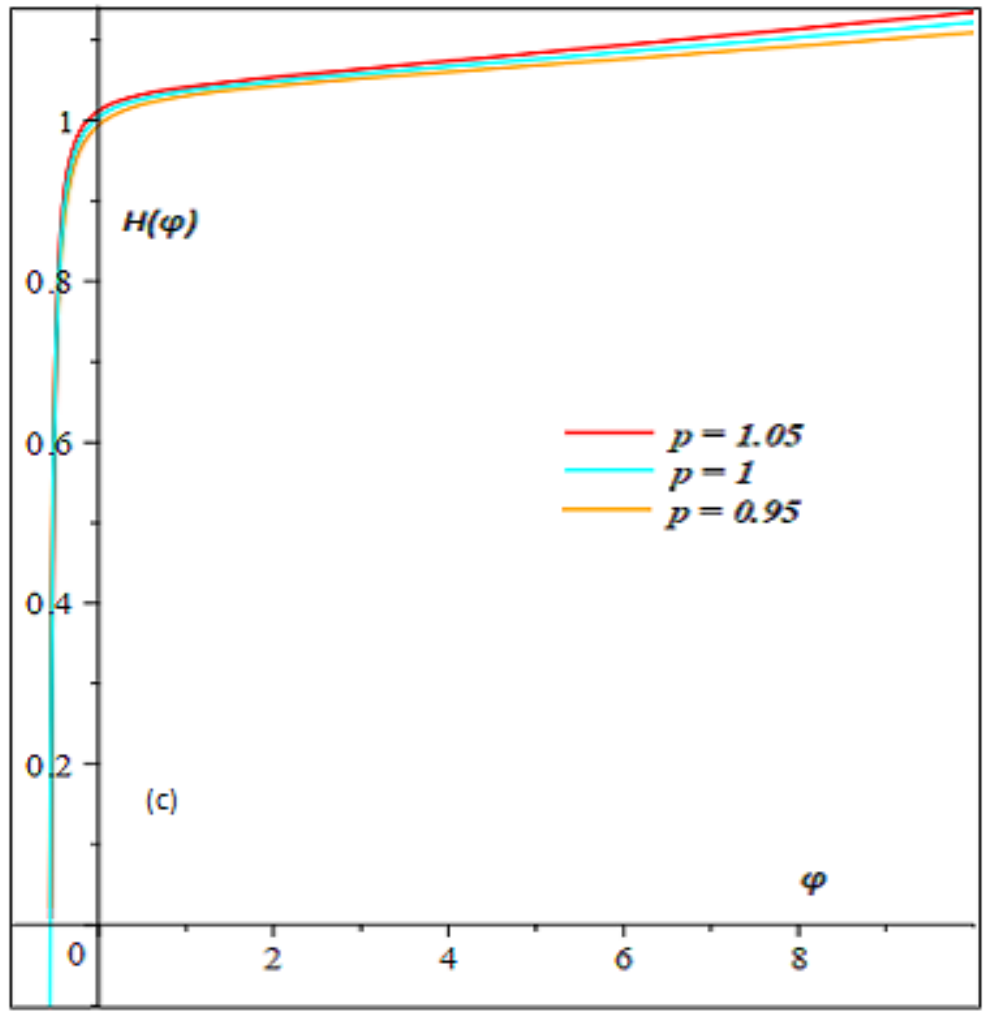}
	\caption{The potential, evolution of scaleron and Hubble parameter  (\ref{38}) plotted versus $\varphi$ for $p=1.05, 1, 0.95$ with $\gamma=-1$, $\beta=-0.02$ and $\lambda=1$.}
	\label{fig2}
\end{figure*}
\subsection{$R^{2p}$ model}
As second case, we study $R^{2p}$ model as a generalization  of Starobinsky $R^{2}$ model which was first introduced in the context of higher derivative theories as
\begin{equation}
f(R)=R+\lambda R^{2p}  
\label{36}
\end{equation}
where $\lambda$ and $p$ are free parameters. In this case, $F$ is obtained by
\begin{equation}
F_{2}=1+2p\lambda\bigg(2\gamma(\beta-3)(\beta-1)e^{-\frac{(1+\beta)\varphi}{\sqrt{6}}}+\frac{2}{3}(\beta-3)(\beta-2)(\beta+3)e^{\frac{2(1-\beta)\varphi}{\sqrt{6}}}\bigg)^{2p-1}
\label{37}
\end{equation}
and now the Hubble parameter (\ref{20}), the potential of scaleron and also its evolution (\ref{21}) can be expressed by
\begin{equation}
\hat{H}_{2}(\varphi)=\bigg(\gamma F_{2}^{-\frac{3}{2}}+F_{2}^{-\frac{\beta}{2}}\bigg),\quad V_{2}(\varphi)=\frac{3-\beta}{3}\bigg(6\gamma F_{2}^{-\frac{(3+\beta)}{2}}+(3+\beta)F_{2}^{-\beta}\bigg),\quad
\frac{d\varphi_{2}}{d\hat{t}}=\frac{2}{\sqrt{6}}\bigg(3\gamma F_{2}^{\frac{-3}{2}}+\beta F_{2}^{-\frac{\beta}{2}}\bigg).
\label{38}
\end{equation}
The Fig. \ref{fig2} shows the potential, the evolution of scaleron and the Hubble parameter (\ref{38}) versus $\varphi$ for different values of $p$ in $R^{2p}$ model with $\gamma=-1$, $\beta=-0.02$ and $\lambda=1$. From the panel (a), we recover the Starobinsky model in the case of $p=1$. For the values $p=1.05$ and $p=0.95$, the potential behaves the same with $p=1$. It seems that a tiny deviation from the Starobinsky model $p=1$ does not provide a remarkable change when inflation is considered in the context of constant-roll approach. This fact also can be understood from the panels (b) and (c) which show the phase diagram of the scaleron and the Hubble parameter of the model, respectively. The slow-roll parameters (\ref{23} - \ref{25}) of the $R^{2p}$ model for the case of $\gamma=0$ are given by
\begin{equation}
\epsilon=\frac{\beta^{2}}{2}\bigg(\frac{F'_{2}}{F_{2}}\bigg)^{2},\quad\eta=-\beta\bigg(\frac{F_{2}''}{F_{2}}-(\beta+1)(\frac{F_{2}'}{F_{2}})^{2}\bigg),\quad\zeta^{2}=\beta^{2}\frac{F_{2}'}{F_{2}^{2}}\bigg(F_{2}'''-3(\beta+1)\frac{F_{2}''F_{2}'}{F_{2}}+(\beta+1)(\beta+2)\frac{F_{2}'^{3}}{F_{2}^{2}}\bigg)
\label{39}    
\end{equation}
and the number of e-folds (\ref{28}) of the $R^{2p}$ model can be calculated as
\begin{equation}
N=\int^{\varphi_{i}}_{\varphi_{f}}\frac{1}{\beta}\frac{F_{2}}{F'_{2}}d\varphi=\frac{-3\bigg(\frac{2}{3}(\beta-2)(\beta-3)(\beta+3)\bigg)^{1-2p}}{4\lambda\beta p(2p-1)^{2}(1-\beta)^{2}}\bigg[\bigg(e^{\frac{2(1-\beta)\varphi_{i}}{\sqrt{6}}}\bigg)^{1-2p}-\bigg(e^{\frac{2(1-\beta)\varphi_{f}}{\sqrt{6}}}\bigg)^{1-2p}\bigg]+\frac{3(\varphi_{i}-\varphi_{f})}{\sqrt{6}\beta(1-\beta)(2p-1)}
\label{40}    
\end{equation}
and by removing the role of scalar field when inflation ends ($\varphi_{f}\ll\varphi_{i}$), $N$ takes the following form
\begin{equation}
N\simeq\frac{-3\bigg(\frac{2}{3}(\beta-2)(\beta-3)(\beta+3)e^{\frac{2(1-\beta)\varphi_{i}}{\sqrt{6}}}\bigg)^{1-2p}}{4\lambda\beta p(2p-1)^{2}(1-\beta)^{2}}.
\label{41}
\end{equation}
By combining the slow-roll parameters (\ref{39}) and the number of e-folds (\ref{41}), the inflationary parameters (\ref{29}) of the $R^{2p}$ model are calculated as shown in (\ref{a4} - \ref{a6}). 
\begin{figure*}[!hbtp]
	\centering
	\includegraphics[width=.28\textwidth,keepaspectratio]{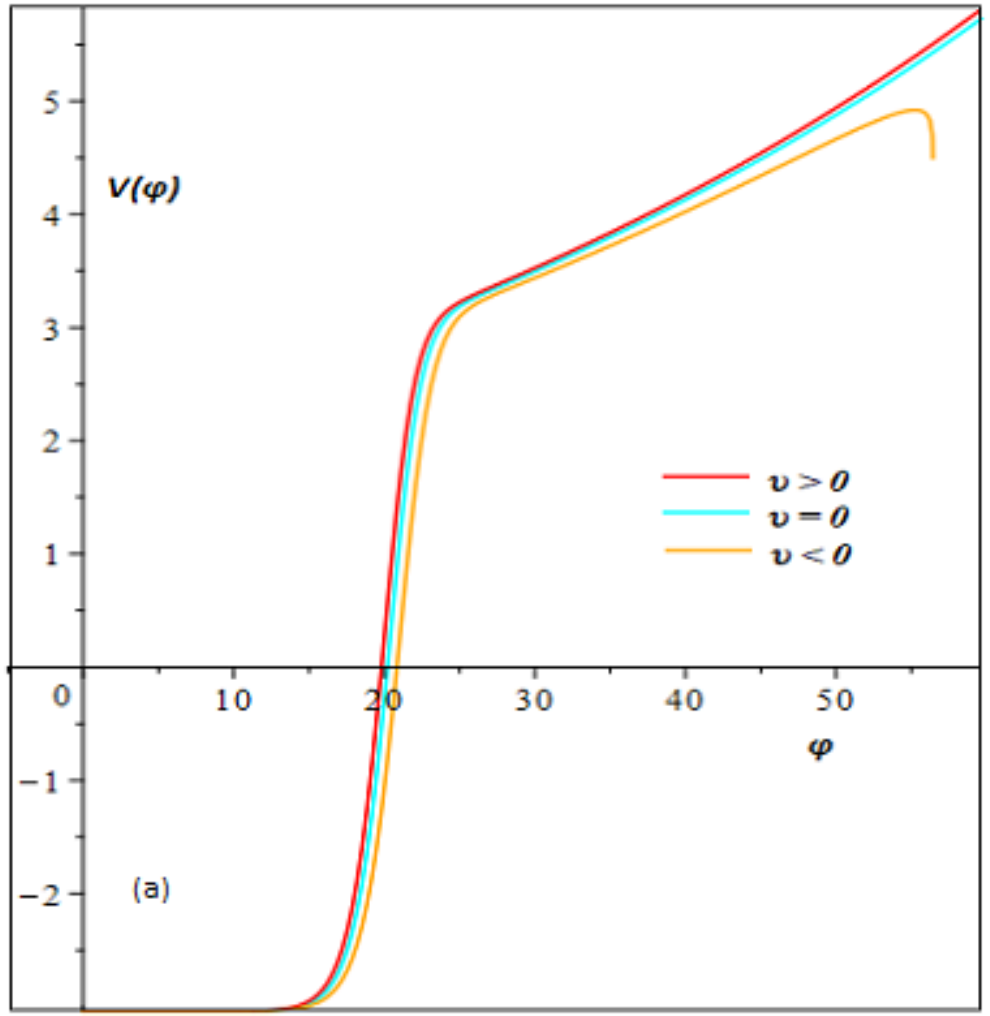}
	\hspace{0.5cm}
	\includegraphics[width=.28\textwidth,keepaspectratio]{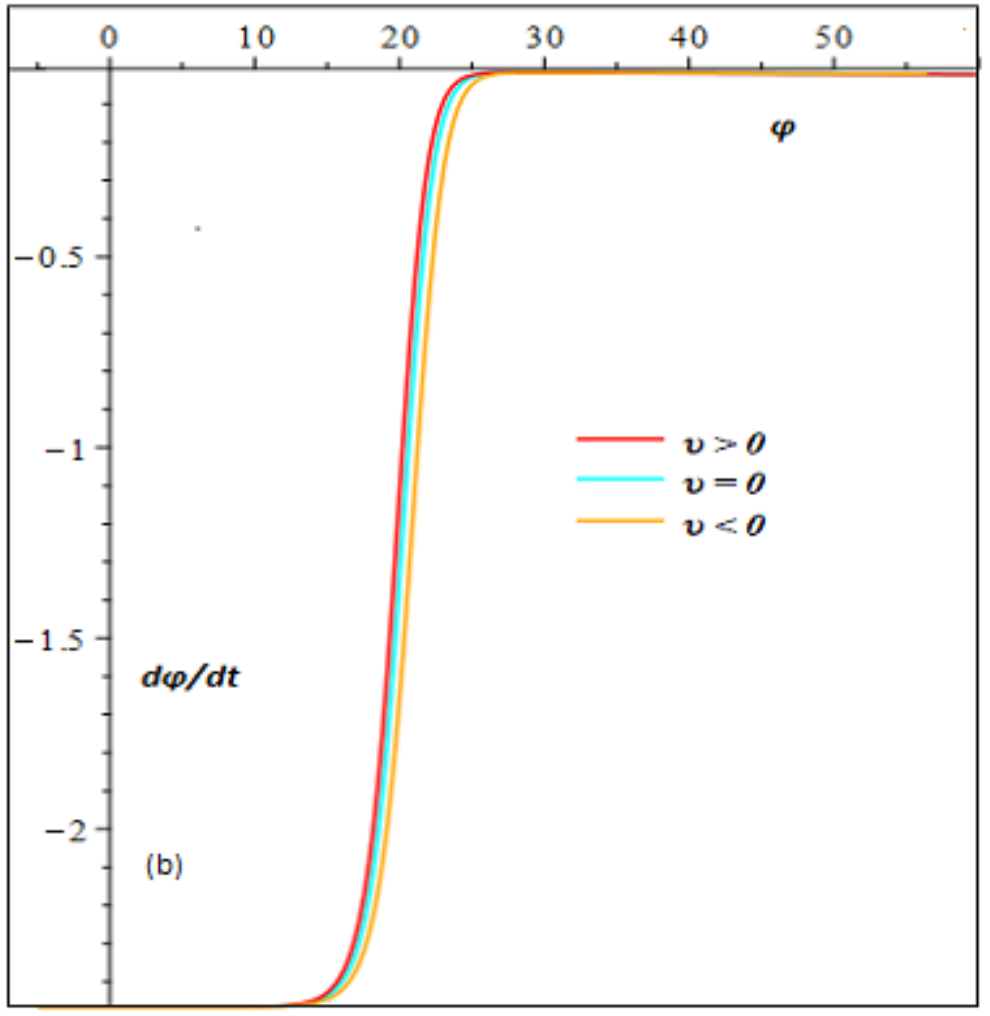}
	\hspace{0.5cm}
	\includegraphics[width=.28\textwidth,keepaspectratio]{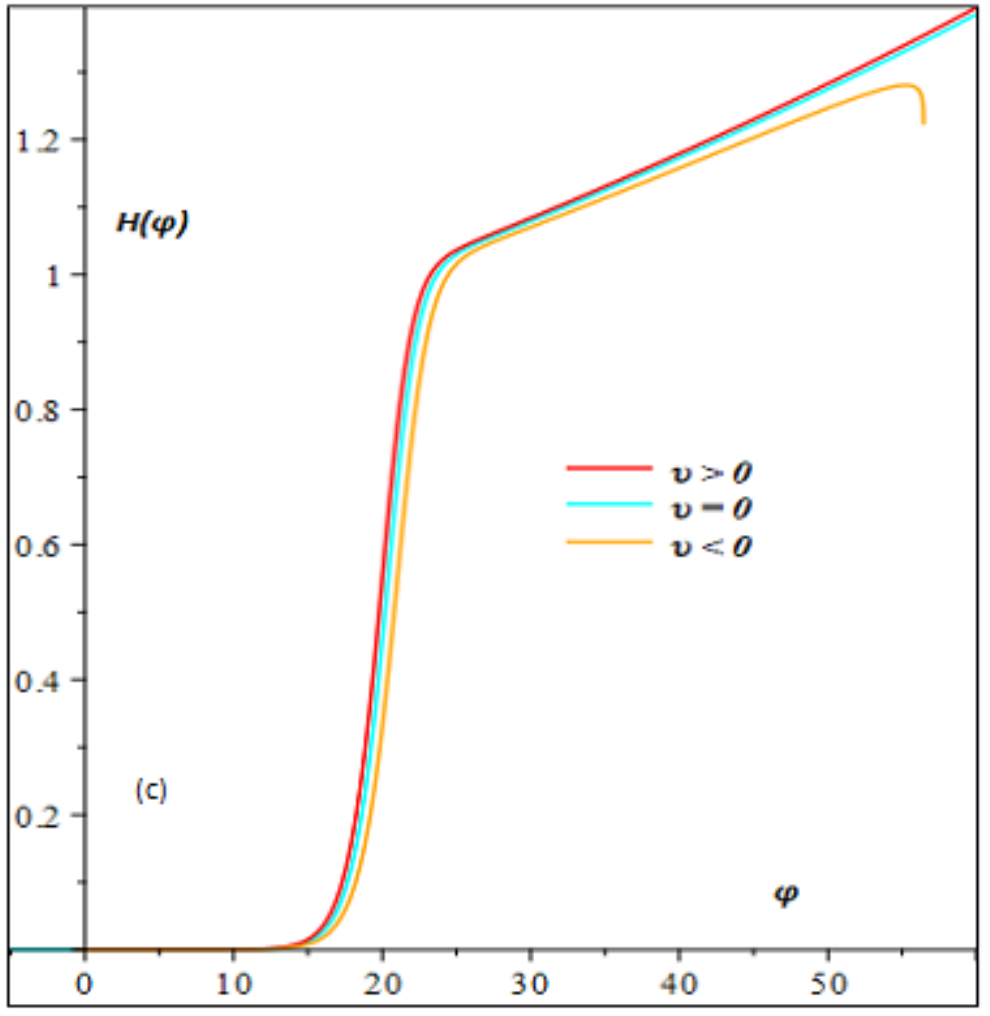}
	\caption{The potential, evolution of scaleron and Hubble parameter  (\ref{44}) plotted versus $\varphi$ for different signs of $\upsilon = 0, |0.02|\lambda$ with $\lambda=1.2\times10^{-9}$, $\gamma=-1$ and $\beta=-0.02$.}
	\label{fig3}
\end{figure*}
\subsection{Logarithmic corrected model}
Let us consider now a logarithmic corrected $f(R)$ model
\begin{equation}
f(R)=R+\lambda R^{2}+\upsilon R^{2}\ln R
\label{42}
\end{equation}
where the phenomenological parameters $\lambda$, $\upsilon$ will be fixed momentarily. In fact, the extra terms come from the leading quantum gravity corrections. See \cite{Birrell,Shapiro} for details.  For the model (\ref{42}), $F$ takes the following form 
\begin{eqnarray}
&\!&\!F_{3}=1+\bigg(2\gamma(\beta-3)(\beta-1)e^{-\frac{(1+\beta)\varphi}{\sqrt{6}}}+\frac{2}{3}(\beta-3)(\beta-2)(\beta+3)e^{\frac{2(1-\beta)\varphi}{\sqrt{6}}}\bigg)\times\nonumber\\&\!&\!
\times\Bigg\{2\lambda+\upsilon+2\upsilon\ln{\bigg(2\gamma(\beta-3)(\beta-1)e^{-\frac{(1+\beta)\varphi}{\sqrt{6}}}+\frac{2}{3}(\beta-3)(\beta-2)(\beta+3)e^{\frac{2(1-\beta)\varphi}{\sqrt{6}}}\bigg)}\Bigg\}\hspace{0.5cm}
\label{43}
\end{eqnarray}
and now the Hubble parameter (\ref{20}), then the potential and also the evolution of scaleron (\ref{21}) are obtained by
\begin{equation}
\hat{H}_{3}(\varphi)=\bigg(\gamma F_{3}^{-\frac{3}{2}}+F_{3}^{-\frac{\beta}{2}}\bigg),\quad V_{3}(\varphi)=\frac{3-\beta}{3}\bigg(6\gamma F_{3}^{-\frac{(3+\beta)}{2}}+(3+\beta)F_{3}^{-\beta}\bigg),\quad
\frac{d\varphi_{3}}{d\hat{t}}=\frac{2}{\sqrt{6}}\bigg(3\gamma F_{3}^{\frac{-3}{2}}+\beta F_{3}^{-\frac{\beta}{2}}\bigg).
\label{44}
\end{equation}
The panel (a) of Fig. \ref{3} exhibits the behaviour of potential and also the phase diagram of the scaleron (\ref{44}) for different signs of $\upsilon$ in the logarithmic $f(R)$ model. The potential behaves almost the same for different signs of $\upsilon$. However, $\upsilon<0$ shows a different behaviour at the end of inflation compared to two other cases. Similar to the previous models, the potential shows a constant rate of rolling which is connected to the standard slow-roll. The slow-roll parameters (\ref{23} - \ref{25}) of the logarithmic corrected model for the case of $\gamma=0$ are obtained as \begin{equation}
\epsilon=\frac{\beta^{2}}{2}\bigg(\frac{F'_{3}}{F_{3}}\bigg)^{2},\quad\eta=-\beta\bigg(\frac{F_{3}''}{F_{3}}-(\beta+1)(\frac{F_{3}'}{F_{3}})^{2}\bigg),\quad\zeta^{2}=\beta^{2}\frac{F_{3}'}{F_{3}^{2}}\bigg(F_{3}'''-3(\beta+1)\frac{F_{3}''F_{3}'}{F_{3}}+(\beta+1)(\beta+2)\frac{F_{3}'^{3}}{F_{3}^{2}}\bigg)
\label{45}    
\end{equation}
and the number of e-folds (\ref{28}) of the logarithmic model discussed in this section is introduced by
\begin{equation}
N=\int^{\varphi_{i}}_{\varphi_{f}}\frac{1}{\beta}\frac{F_{3}}{F'_{3}}d\varphi=\frac{6e^{\frac{2\lambda+3\upsilon}{2\upsilon}}}{8\beta(1-\beta)^{2}\upsilon}\mbox{Ei}\bigg(-\frac{3\upsilon+2\lambda+2\upsilon\ln\bigg(\frac{2}{3}(\beta-1)(\beta-3)(\beta+3)e^{\frac{2(1-\beta)\varphi}{\sqrt{6}}}\bigg)}{2\upsilon}\bigg)\bigg|^{\varphi_{i}}_{\varphi_{f}}
\label{46}    
\end{equation}
where \mbox{Ei} is the exponential integral that can be expressed by the Puiseux series 
\begin{equation}
\mbox{Ei}(x)=\rho+\ln(x)+x+\frac{x^{2}}{4}+\frac{x^{3}}{18}+\frac{x^{4}}{96}+...\quad\quad with\quad\quad e^{2\rho}=3.17221895...
\label{47} 
\end{equation}
along the positive real axis. Finally, by keeping the logarithmic term and also neglecting the value of scaleron at the end of inflation ($\varphi_{f}\ll\varphi_{i}$), $N$ is written as 
\begin{eqnarray}
N\simeq\frac{6e^{\frac{2\lambda+3\upsilon}{2\upsilon}}}{8\beta(1-\beta)^{2}\upsilon}\ln\Bigg\{-\frac{3\upsilon+2\lambda+2\upsilon\ln\Big(\frac{2}{3}(\beta-1)(\beta-3)(\beta+3)e^{\frac{2(1-\beta)\varphi_{i}}{\sqrt{6}}}\Big)}{2\upsilon}\Bigg\}.
\label{48}
\end{eqnarray}
\begin{figure*}[!hbtp]
	\centering
	\includegraphics[width=0.325\textwidth,keepaspectratio]{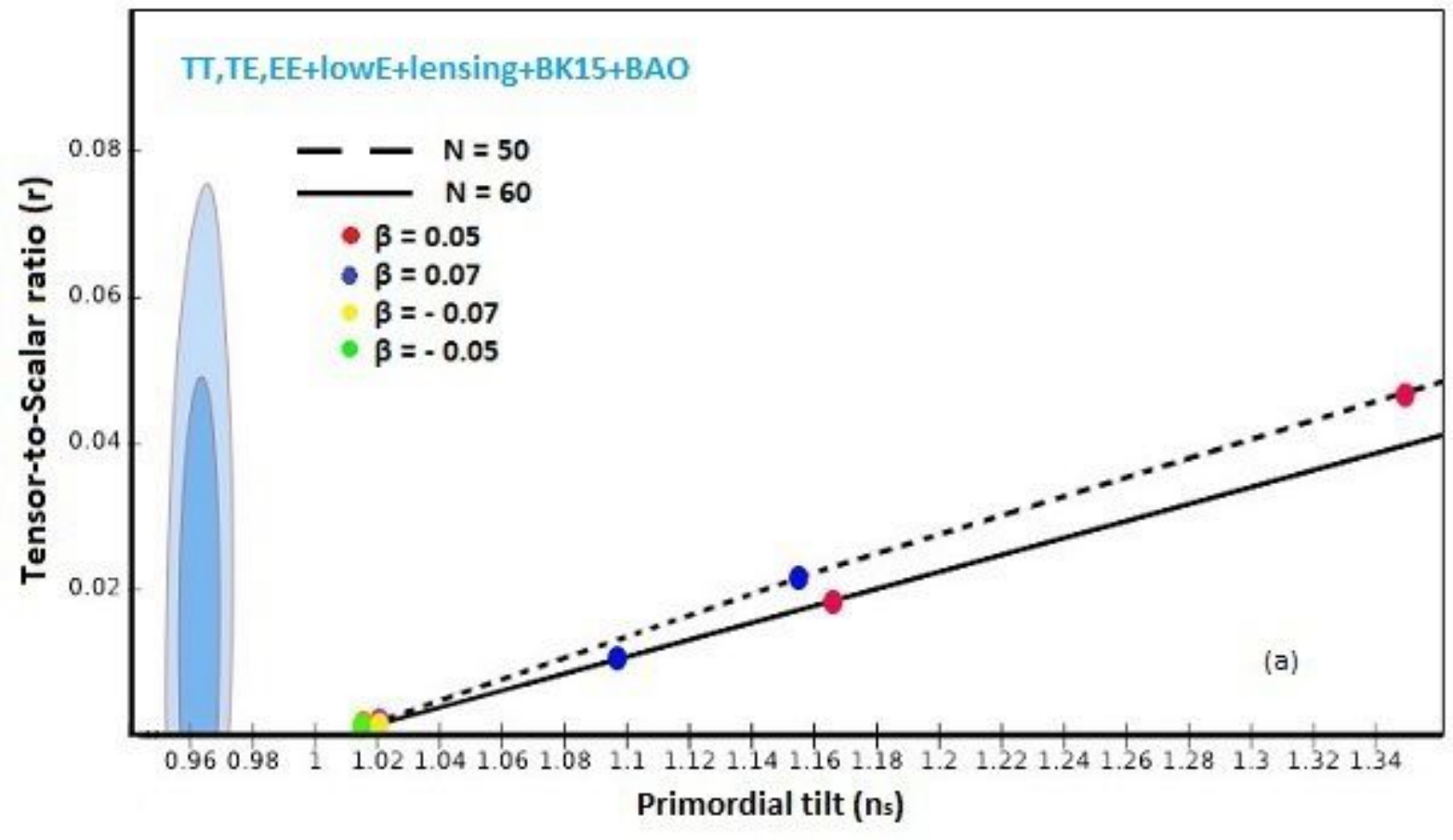}
    \includegraphics[width=0.325\textwidth,keepaspectratio]{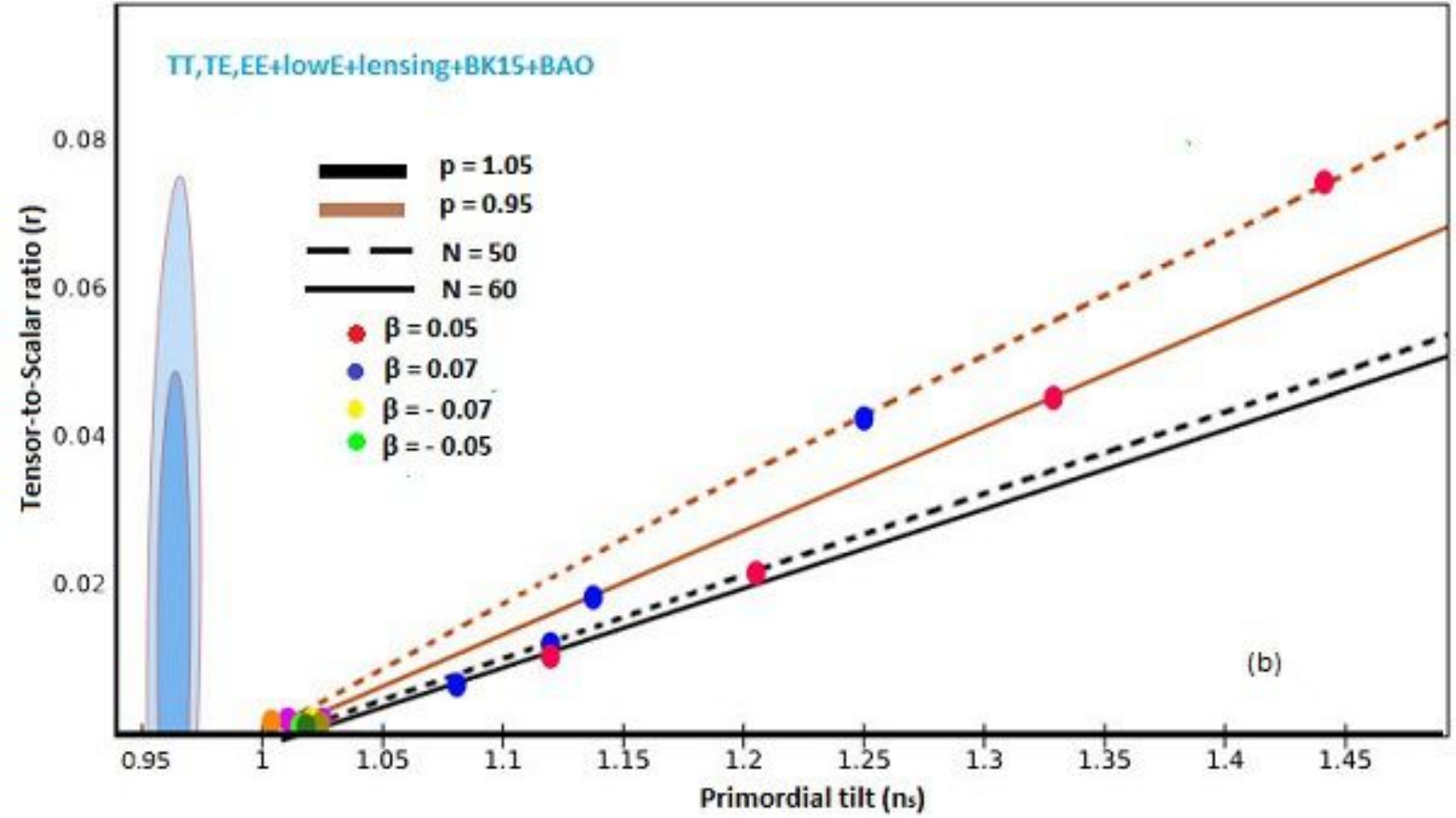}
    \includegraphics[width=0.315\textwidth,keepaspectratio]{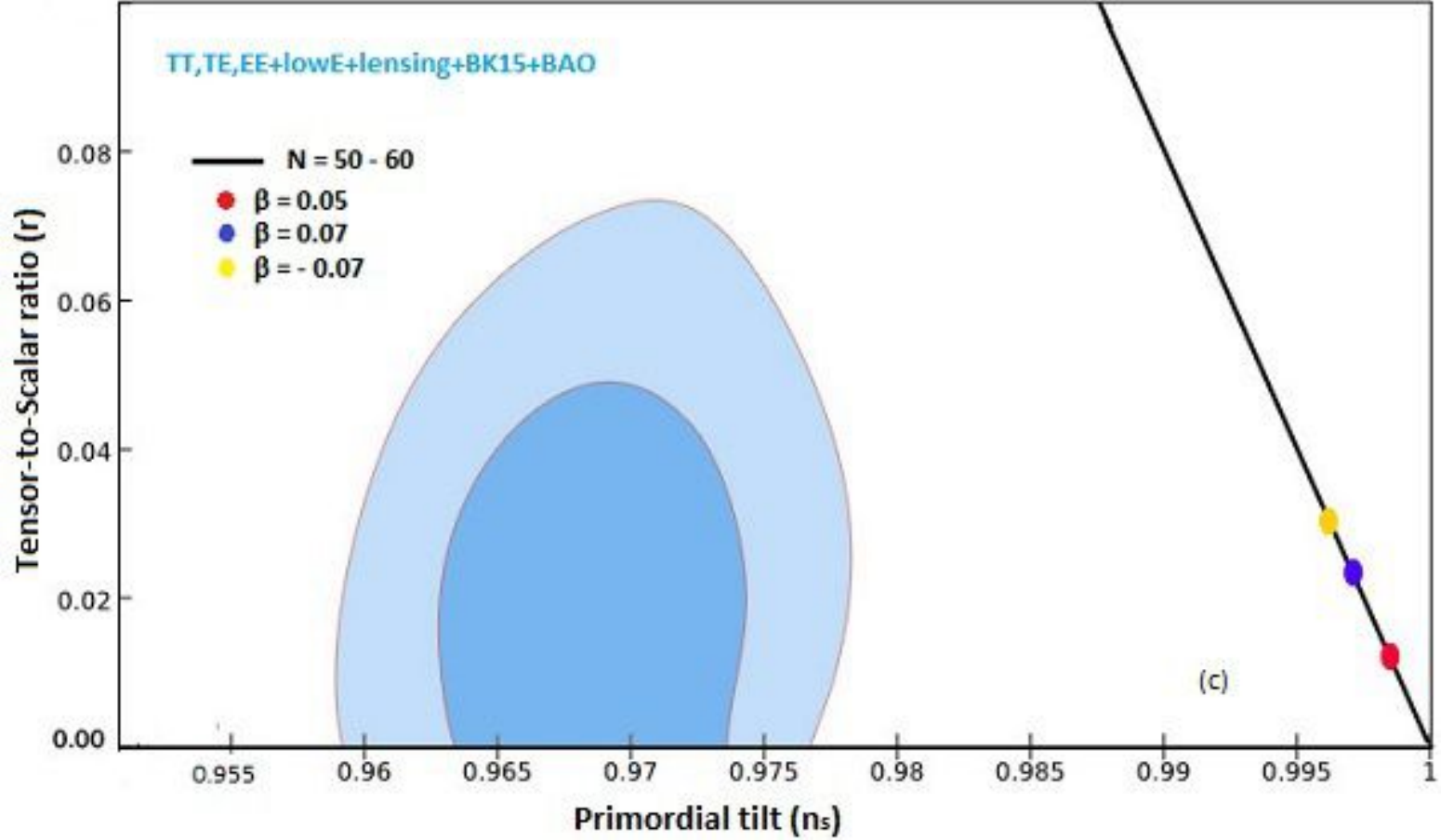}
	\caption{The marginalized joint 68\% and 95\% CL regions for $n_{s}$ and $r$ at $k = 0.002$ Mpc$^{-1}$ from Planck 2018 combining with BK15+BAO data \cite{cmb} and the $n_{s}-r$ constraints on different $f(R)$ inflationary models described in the context of the constant-roll idea. The dashed and solid lines represent $N=50$ and $N=60$, respectively. The results are obtained for some allowed values of $\beta$ when $\gamma=0$ for Starobinsky $R^{2}$ model (a), $\gamma=0$ and $\lambda=1$ for $R^{2p}$ model (b), $\gamma=0$ and $\upsilon=0.02\lambda$ for logarithmic corrected model (c).}
	\label{fig4}
\end{figure*}
By considering the slow-roll parameters (\ref{45}) and the number of e-folds (\ref{48}), the inflationary parameters (\ref{29}) of the logarithmic corrected model are obtained as shown in (\ref{a8} - \ref{a10}).
\section{Comparison with observations}
In this section, we compare the obtained results of $f(R)$ models considered in the paper with the inflationary observations coming from temperature and polarization anisotropies of CMB. 

In Figure \ref{fig4}, we present the $n_{s} - r$ constraints coming from the marginalized joint 68\% and 95\% CL regions of the Planck 2018 in combination with BK15 and BAO data on the $f(R)$ inflationary models  \textit{i.e.} the Starbionsky $R^{2}$ model, the $R^{2p}$ model and the logarithmic corrected model described in the context of the constant-roll approach. The panels are drawn for some allowed values of $\beta$ in the cases $N=50$ (dashed line) and $N=60$ (solid line). Panel (a) is belonged to the Starobisnky $R^{2}$ model (\ref{30}) when $\gamma=0$ in the Eq. (\ref{20}). As we can see, the obtained value of tensor-to-scalar ratio $r$ for the allowed values of $\beta$ is in good agreement with the Planck 2018 \cite{cmb} constraint $r<0.064$, in particular, for the cases of $\beta=0.05$ and $\beta=0.07$ which show $r=0.047341$ and $r=0.021806$, respectively. The negative cases of $\beta$ predict a tiny tensor-to-scalar ratio $r=\mathcal{O}(10^{-3})$ which is beyond the reach of the current observations. Despite the mentioned success, the Starobinsky $R^{2}$ model does not show a scale-invariant spectrum since the value of the spectral index $n_{s}$ for the allowed values of $\beta$ is bigger than the unit. Panel (b) is dedicated to show the Planck constraints on the $R^{2p}$ model (\ref{36}) when $\gamma=0$ in the Eq. (\ref{20}) and $\lambda=1$ for two interesting cases of $p=1.05$ and $p=0.95$. The panel shows that by considering a tiny variation from the original Starobinsky model ($p=1$), the obtained values of $r$ are still in good agreement with the Planck constraint in exception the case of $\beta=0.05$ in $p=0.95$ which presents a disfavoured value $r\simeq0.077$. Also, the negative values of $\beta$ reveal a tiny tensor-to-scalar ratio $r=\mathcal{O}(10^{-3})$ for both cases of $p$. Similar to the Starobinsky model, the $R^{2p}$ model shows observationally disagreeable values of the spectral index $n_{s}>1$ for both cases of $p$. In overall, by taking a look at the obtained values of $r$ and $n_{s}$, we find that the case of $p=0.95$ is more desirable than $p=1.05$ since the values of $r$ are closer to the upper limit ($r<0.064$) of the current observations. In contrast, by focusing on the $n_{s}$, one can find that the deviation from $n_{s}=1$ in the case of $p=1.05$ is remarkably less than the case of $p=0.95$. Finally, panel (c) is corresponded to the interesting case of the logarithmic corrected model (\ref{42}) when $\gamma=0$ in the Eq. (\ref{20}) and $\upsilon=0.02\lambda$. From the panel, we can see that the obtained values of $r$ for the allowed values of $\beta$ are compatible with the observational constraint coming from Planck 2018. This is valid for both positive and negative values of $\beta=0.05$, $0.07$ and $-0.07$ which show $r=0.012266$, $0.023041$ and $0.0305$, respectively. Concerning the spectral index, the panel tells us that the logarithmic corrected model shows a scale-invariant spectrum ($n_{s}<1$) but still situated in a disfavorate region $0.996\leq n_{s}\leq 0.999$ which is far from the Planck constraint $n_{s}=0.9649\pm0.0042$. In comparison with the previous models, here the negative values of $\beta$ predict more desirable results of $r$ and $n_{s}$ than the positive values of $\beta$. 

\begin{figure*}[!hbtp]
	\centering
	\includegraphics[width=.245\textwidth,keepaspectratio]{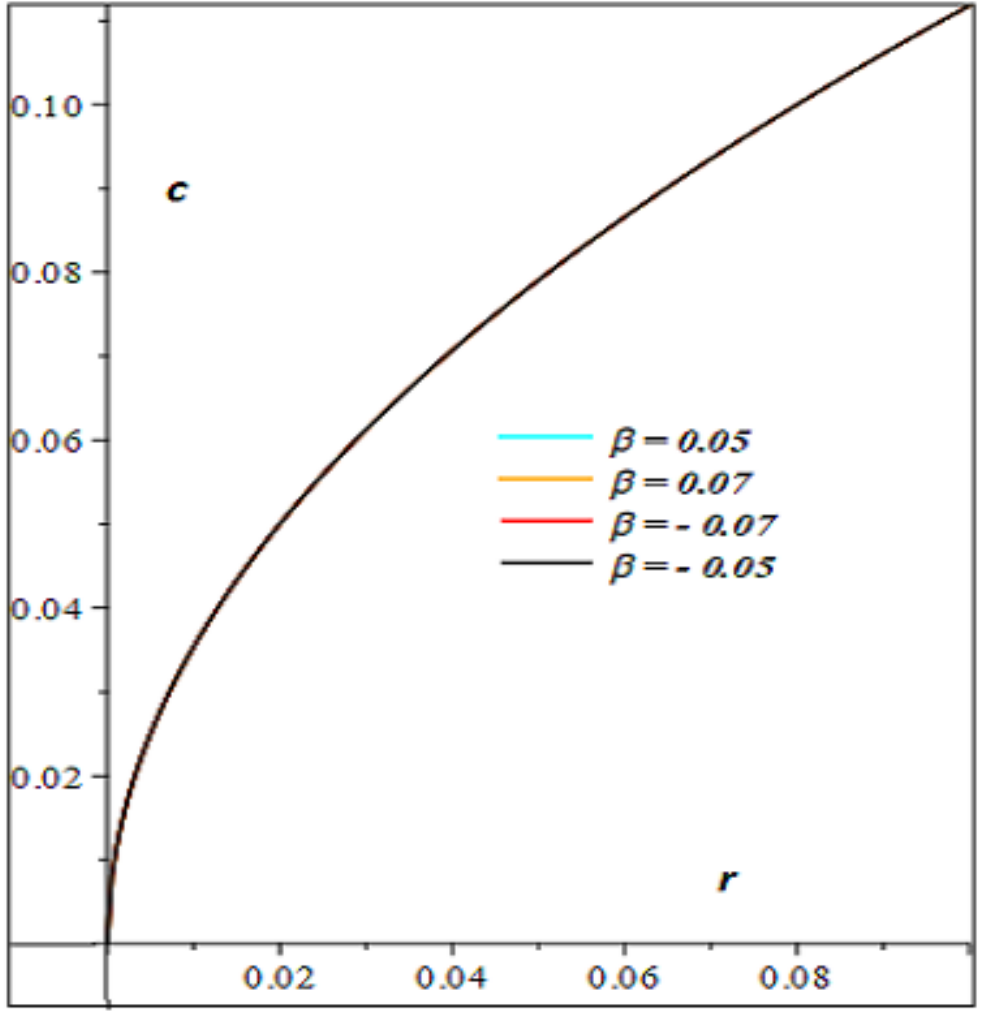}
	\includegraphics[width=.245\textwidth,keepaspectratio]{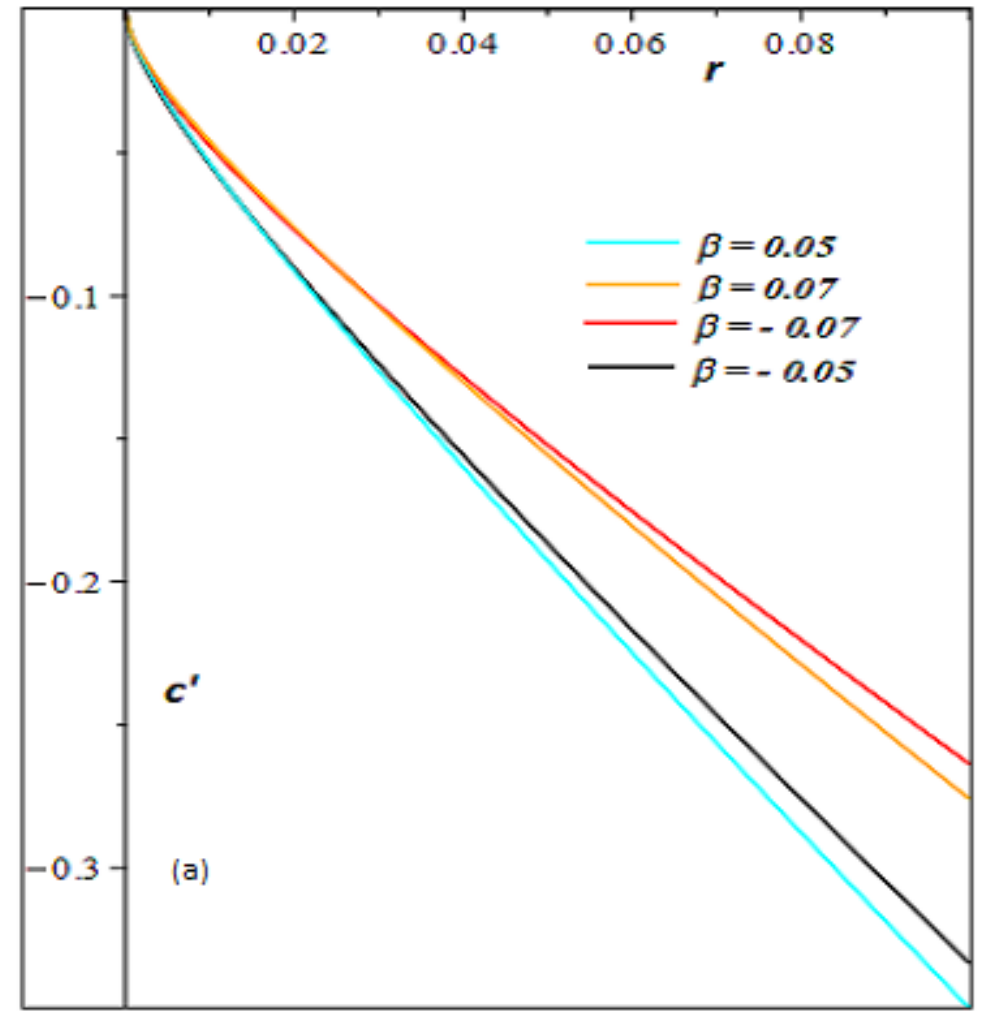}
	\includegraphics[width=.245\textwidth,keepaspectratio]{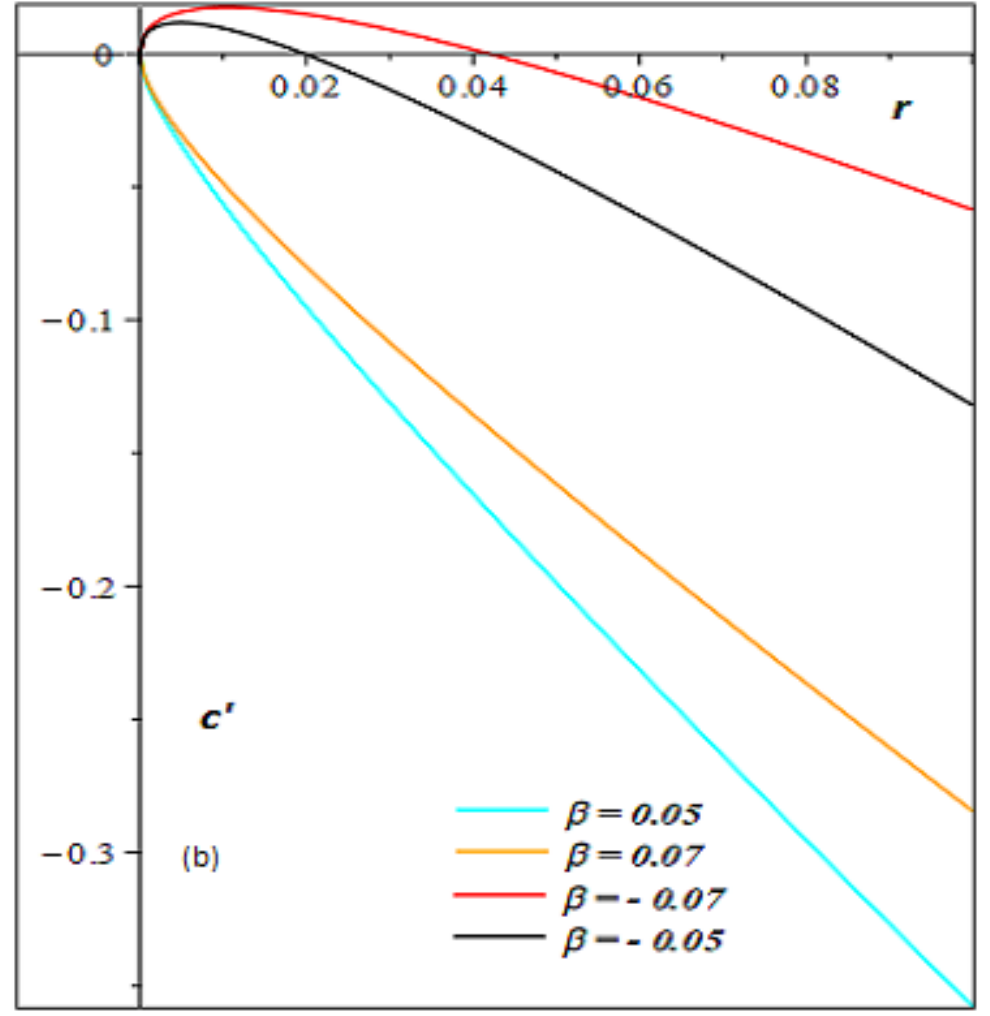}
	\includegraphics[width=.245\textwidth,keepaspectratio]{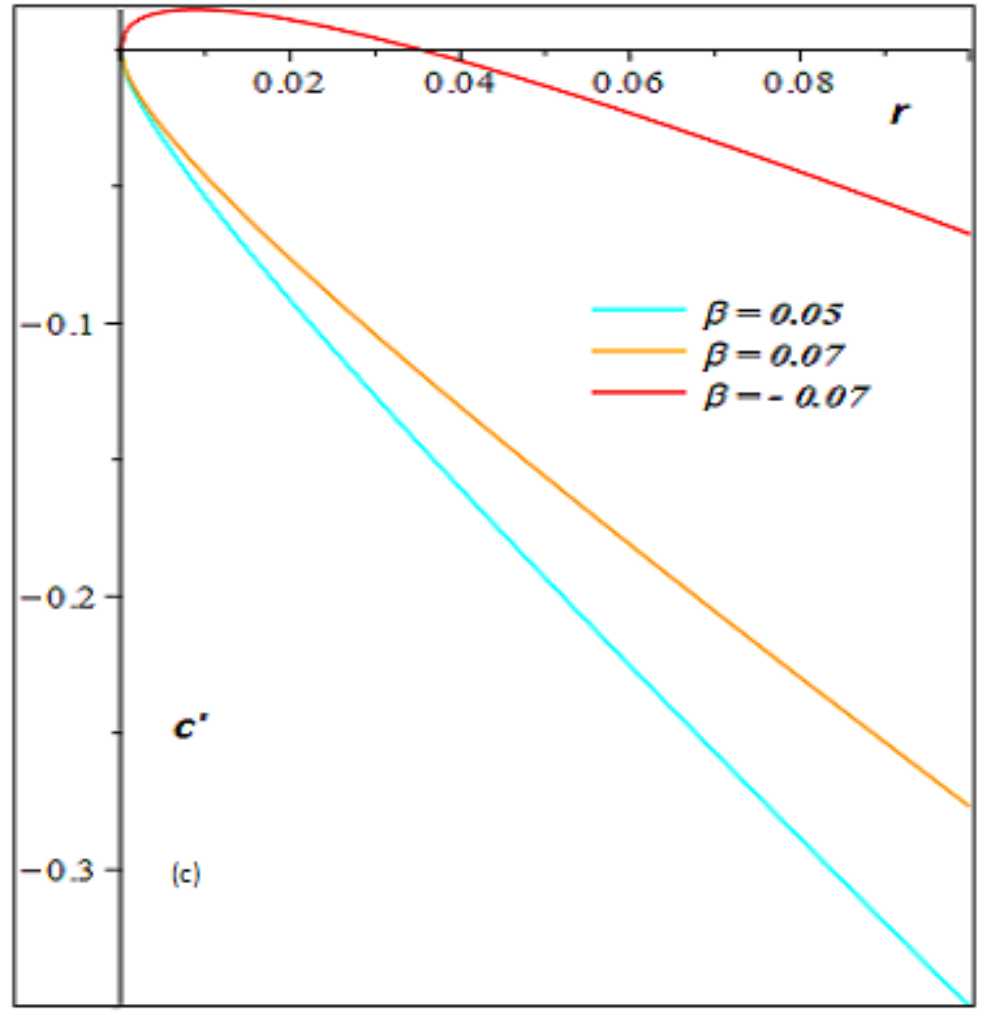}
	\caption{The behaviour of the swampland parameters $c$ and $c'$ (\ref{26}) and (\ref{27}) versus the tensor-to-scalar ratio $r$ for different $f(R)$ inflationary models described in the context of the constant-roll approach. The results are obtained for some allowed values of $\beta$ when $\gamma=0$ for the Starobinsky $R^{2}$ model (a), $\gamma=0$ and $\lambda=1$ for the $R^{2p}$ model (b), $\gamma=0$ and $\upsilon=0.02\lambda$ for the logarithmic corrected model (c).}
	\label{fig5}
\end{figure*}

On the other hand, we can study the issue from the viewpoint of the swampland criteria (\ref{26}) and (\ref{27}). In Table \ref{tab1}, we present the values of the swampland parameters $c$ and $c'$ for all three $f(R)$ constant-roll models. The results are obtained for some allowed values of $\beta$ with $N$ between 50 and 60 when $\gamma=0$ for the Starobinsky $R^{2}$ model, $\gamma=0$ and $\lambda=1$ for the $R^{2p}$ model and $\gamma=0$ and $\upsilon=0.02\lambda$ for the logarithmic corrected model. Also, we investigate the behaviour of the swampland parameters $c$ and $c'$ versus the tensor-to-scalar ratio $r$ for the considered models in Figure \ref{fig5}. Now, let's review the swampland conditions for our $f(R)$ constant-roll inflationary models.
\begin{table}
\begin{center}
\begin{tabular}{|c|c|c|c|c|}
  \hline
   Model& $\beta$ & $N$ & $c\leq$& $c'\leq$ \\
  \hline
   & $0.05$ & $50-60$ & $0.077065$ & $-0.18394$ \\
  \cline{2-5}
   Starobinsky $R^{2}$ & $0.07$ & $50-60$ & $0.052303$ & $-0.081309$  \\ \cline{2-5}
   & $-0.07$ & $50-60$ & $0.016687$ & $-0.010866$  \\ \cline{2-5}
   & $-0.05$ & $50-60$ & $0.015134$ & $-0.0086153$  \\
  \hline
  & $0.05$ & $50-60$ & $0.052024$ & $-0.101226$ \\
  \cline{2-5}
   $R^{2p}(p=1.05)$ & $0.07$ & $50-60$ & $0.04055$ & $-0.059005$  \\ \cline{2-5}
   & $-0.07$ & $50-60$ & $0.015865$ & $-0.011918$  \\ \cline{2-5}
   & $-0.05$ & $50-60$ & $0.014602$ & $-0.0097295$  \\ 
  \hline 
  & $0.05$ & $50-60$ & $0.15984$ & $-0.64812$ \\
  \cline{2-5}
  $R^{2p}(p=0.95)$ & $0.07$ & $50-60$ & $0.075377$ & $-0.13836$  \\ \cline{2-5}
    & $-0.07$ & $50-60$ & $0.017365$ & $-0.0096579$  \\
  \cline{2-5}
  & $-0.05$ & $50-60$ & $0.015475$ & $-0.0073956$  \\
  \hline
  & $0.05$ & $50-60$ & $0.039158$ & $-0.0015361$ \\
  \cline{2-5}
   Logarithmic corrected & $0.07$ & $50-60$ & $0.053666$ & $-0.0028838$  \\ \cline{2-5}
   & $-0.07$ & $50-60$ & $0.061746$ & $-0.0038075$  \\
  \hline
\end{tabular}
  \caption{The values of the swampland parameters $c$ and $c'$ coming from the swampland conditions (\ref{26}) and (\ref{27}) for different $f(R)$ inflationary models described in the context of the constant-roll approach. The results are obtained for some allowed values of $\beta$ with $N=50$ when $\gamma=0$ for the Starobinsky $R^{2}$ model, $\gamma=0$ and $\lambda=1$ for the $R^{2p}$ model, $\gamma=0$ and $\upsilon=0.02\lambda$ for the logarithmic corrected model.}
  \label{tab1}
\end{center}
\end{table}
In the Starobinsky $R^{2}$ model, we find that the swampland condition of the parameter $c$ shown in the table leads to some desirable values of the tensor-to-scalar ratio $r$, in particular, for the positive cases of $\beta$ and this coincidence is distorted for tiny values of $c$ since the corresponding values of $r$ are not situated in the range of our current observations (see the first panel of Figure \ref{5}). For another swampland parameter $c'$, the table tells us in the case of $\beta=0.05$, $c'$ takes the value $-0.18394$ which is connected to an  observationally acceptable value $r\simeq0.05$. By considering the swampland condition $c'$, the value of $r$ approaches the observational upper limit and then reaches disfavoured regions of $r$ by crossing from the limit. For other values of $\beta$, we find that the values of $c'$ are so tiny and lead to the small values of $r$. By setting the swampland condition $c'$, the situation becomes better since the corresponding value of $r$ approaches the upper limit (see panel (a) of Figure \ref{5}). In the $R^{2p}$ model with $p=1.05$, the situation of $c$ is almost similar to the Starobinsky $R^{2}$ model while the parameter $c'$ reveals some interesting features. The values of $\beta=0.05,0.07,-0.05$ show $r\simeq0.02,0.01,0.02$ which are close to the lower limit of the observations and by considering the swampland condition, $r$ approaches the upper limit. For $\beta=-0.07$, we find $r\simeq0.7$ which is situated beyond the upper limit (see panel (b) of Figure \ref{5}). In the case of $p=0.95$, the parameter $c$ behaves analogous to the Starobinsky $R^{2}$ model in exception the case of $\beta=0.05$ which leads to $r$ bigger than the upper limit ($r<0.064$). Also, for $\beta=0.07,-0.07,-0.05$, the parameter $c'$ shows the values of $r\simeq0.03,0.05,0.02$ so that by setting the swampland condition, they approach the upper limit. Note that for $\beta=0.05$, the parameter $c'$ predicts a very large $r$ which is not compatible with the observations. In the logarithmic corrected model, the obtained values of $c$ present $r$ below the upper limit and again by considering the swampland condition, they approach the lower limit. For the parameter $c'$, $\beta=-0.07$ shows $r\simeq0.04$ while $\beta=0.05,0.07$ predict tiny $r$ (see panel (c) of Figure \ref{5}). 
\section{Discussion and conclusions}
In the present work, we  focused on some  $f(R)$ inflationary models \textit{i.e.} the Starobinsky $R^{2}$, the $R^{2p}$ and the logarithmic corrected models introduced in the context of the constant-roll  where inflaton rolls down with a constant rate as $\ddot{\varphi}=\beta H\dot{\varphi}$. We have investigated the inflationary dynamics for the  three $f(R)$ models in  presence of  constant-roll condition. Specifically, we  calculated the spectral parameters of the models \textit{i.e.} the spectral index, its running and the tensor-to-scalar ratio. Then, we  compared the obtained results with Planck 2018 data combined with BK15+BAO data.
Results can be summarized as follows :

\begin{itemize}

    \item We  studied the Starobinsky $R^{2}$ model for $\gamma=0$ and  found that,  for  positive values of $\beta$, $r$ is in good agreement with  observations while, for   negative values of $\beta$, it shows a tiny $r=\mathcal{O}(10^{-3})$. Furthermore, the Starobinsky model does not present a scale-invariant spectrum since the obtained values of the spectral index $n_{s}$ are bigger than the unit. 

    \item In the $R^{2p}$ model, we  investigated the model for the two cases  $p=1.05$ and $p=0.95$ when $\gamma=0$ and $\lambda=1$. We  found that the obtained values of $r$ are compatible with the observational constraint on $r$, except  the case  $\beta=0.05$ in $p=0.95$ which predicts $r\simeq0.077$ bigger than the upper limit ($r<0.064$). Also, a negative $\beta$ presents a tiny $r=\mathcal{O}(10^{-3})$ for both cases of $p$. Moreover, the model shows the spectral index $n_{s}>1$ which is not in good agreement with the Planck constraint. In summary, we  found the case of $p=0.95$  more desirable than $p=1.05$ since the values of $r$ are closer to the upper limit ($r<0.064$). In contrast, the deviation from $n_{s}=1$ in the case of $p=1.05$ is less than the case of $p=0.95$.

    \item As a modified form of the Starobinsky model, we have considered the logarithmic corrected model when $\gamma=0$ and $\upsilon=0.02\lambda$. We have found that the obtained values of $r$ for positive and negative values of $\beta$ are in good agreement with the observations. Also, the model predicts the spectral index $n_{s}<1$ but it is still situated in a disfavoured region $0.996<n_{s}<0.999$.
\end{itemize}

According to the above results, one can conclude  that although all three  constant-roll $ f(R)$ inflationary models \textit{i.e.} the Starobinsky $R^{2}$ model, the $R^{2p}$ model and the logarithmic corrected model are observationally compatible with the values of $r$, they do not show acceptable values for the spectral index $n_{s}$. Consequently, the considered  constant-roll $f(R)$ inflationary models are  disfavored models when $\gamma=0$ in the Eq. (\ref{20}). This result can be referred to Ref. \cite{Motohashi9} where the authors studied the behavior of  constant-roll $f(R)$ inflation in a general approach. They found that the model can lead to an inflationary solution only for $-0.1\leq\beta\leq0$ when $\gamma=-1$ in the Eq. (\ref{20}). 

Furthermore, we have studied the models from the viewpoint of the Weak Gravity Conjecture using the swampland criteria. By setting the condition of $c$, we  found that the scalar-to-tensor ratio $r$ starts from an observationally acceptable value and approaches the lower limit. Finally, by crossing from the lower limit, it tilts to the disfavored areas of $r$. Also, by setting the condition of $c'$, the value of $r$ begins from a desirable value and finally, by crossing from the upper limit, it shifts to the disagreeable regions. Hence, we can conclude that our $f(R)$ inflationary models introduced in the context of the constant-roll idea are not fully compatible with the swampland criteria.
\bibliographystyle{ieeetr}
\bibliography{biblo}
\appendix
\section{The spectral parameters of $f(R)$ models}
\subsection{Starobinsky $R^{2}$ model}
\noindent The spectral index of the Starobinsky $R^{2}$ model:
\begin{eqnarray}
&\!&\!n_{s}=\frac{1}{\Big(2N\beta^{3}-4N\beta^{2}+2N\beta-3\Big)^{2}}\Bigg\{(4N^{2}+8N)\beta^{6}-(16N^{2}+32N)\beta^{5}+(24N^{2}+48N-6)\beta^{4}+\nonumber\\&\!&\!
+(-16N^{2}-44N+12)\beta^{3}+(4N^{2}+32N-6)\beta^{2}-12\beta N+9\Bigg\}.
\label{a1}    
\end{eqnarray}
The tensor-to-scalar ratio of the Starobinsky $R^{2}$ model:
\begin{equation}
r=\frac{48(\beta-1)^{2}\beta^{2}}{\Big(2N\beta^{3}-4N\beta^{2}+2N\beta-3\Big)^{2}}.
\label{a2}    
\end{equation}
The running spectral index of the Starobinsky $R^{2}$ model:
\begin{equation}
\alpha_{s}=-\frac{32N\Big(N\beta^{2}-\beta N-\frac{3}{2}\Big)\beta^{3}(\beta-1)^{7}}{\Big(2N\beta^{3}-4N\beta^{2}+2N\beta-3\Big)^{4}}.
\label{a3}    
\end{equation}
\subsection{$R^{2p}$ model}
\noindent The spectral index of the $R^{2p}$ model:
\begin{eqnarray}
&\!&\!n_{s}=\frac{1}{\Bigg(3p\lambda4^{p}9^{-p}\mathcal{W}^{-1}\Big((\beta-2)(\beta^{2}-9)\Big)^{2p}+\beta^{3}-2\beta^{2}-9\beta+18\Bigg)^{2}}\Bigg\{-3p^{2}4^{2p}9^{-2p}\lambda^{2}\mathcal{W}^{-2}\times\nonumber\\&\!&\!
\times\bigg(-3+(p-\frac{1}{2})^{2}\Big(16\beta^{3}-32\beta^{2}+16\beta\Big)\bigg)\Big((\beta-2)(\beta^{2}-9)\Big)^{4p}-16(\beta-3)(\beta-2)4^{p}9^{-p}\lambda p(\beta+3)\mathcal{W}^{-1}\times\nonumber\\&\!&\!
\times\bigg(-\frac{3}{8}+(p-\frac{1}{2})^{2}\Big(\beta^{3}-2\beta^{2}+\beta\Big)\bigg)\Big((\beta-2)(\beta^{2}-9)\Big)^{2p}-24(\beta-2)\Bigg((p-\frac{1}{2})^{2}\beta\mathcal{W}^{-2}\lambda^{2}p^{2}81^{-p}16^{p}\times\nonumber\\&\!&\!
\times(\beta-1)^{2}\Big((\beta-2)(\beta^{2}-9)\Big)^{4p}-\frac{(\beta-2)(\beta-3)^{2}(\beta+3)^{2}}{24}\Bigg)\Bigg\}.
\label{a4}  
\end{eqnarray}
The tensor-to-scalar ratio of the $R^{2p}$ model:
\begin{equation}
r=\frac{192(p-\frac{1}{2})^{2}\beta^{2}\Big((\beta-2)(\beta^{2}-9)\Big)^{4p}\mathcal{W}^{-2}\lambda^{2}p^{2}81^{-p}(\beta-1)^{2}16^{p}}{\Bigg(3p\lambda 4^{p}9^{-p}\mathcal{W}^{-1}\Big((\beta-2)(\beta^{2}-9)\Big)^{2p}+\beta^{3}-2\beta^{2}-9\beta+18\Bigg)^{2}}.
\label{a5}    
\end{equation}
The running spectral index of the $R^{2p}$ model:
\begin{eqnarray}
&\!&\!\alpha_{s}=\frac{1}{\Bigg(-3p\lambda 4^{p}\mathcal{W}^{-1}\Big((\beta-2)(\beta^{2}-9)\Big)^{2p}+3^{(2+2p)}\beta-(\beta^{3}-2\beta^{2}+18)9^{p}\Bigg)^{4}}\Bigg\{-128(p-\frac{1}{2})^{4}\beta^{2}p^{2}\times\nonumber\\&\!&\!
\times(\beta-2)(\beta-3)(\beta+3)(\beta-1)^{4}\lambda^{2}\Bigg(\mathcal{W}^{-2}1296^{p}(\beta-2)(\beta-3)(\beta+3)\Big((\beta-2)(\beta^{2}-9)\Big)^{4p}+3p64^{p}9^{p}\lambda\times\nonumber\\&\!&\!
\times\mathcal{W}^{-3}(\beta-1)\Big((\beta-2)(\beta^{2}-9)\Big)^{6p}\Bigg)\Bigg\}
\label{a6}    
\end{eqnarray}
where
\begin{equation}
\mathcal{W}=-\frac{2\beta p\lambda9^{-p}(-1+b)^{2}(2p-1)^{2}N4^{p}(\beta^{3}-2\beta^{2}-9\beta+18)^{2p}}{(\beta-2)(\beta^{2}-9)}.
\label{a7}   
\end{equation}
\subsection{Logarithmic corrected mode}
\noindent The spectral index of the Logarithmic corrected model:
\begin{eqnarray}
&\!&\!n_{s}=\frac{1}{12\Bigg(e^{\mathcal{Z}}\upsilon(\beta-2)\ln\Big({\frac{(\beta-2)e^{\mathcal{Z}}}{\beta-1}}\Big) +(\lambda+\frac{\upsilon}{2})(\beta-2)e^{\mathcal{Z}}+\frac{(\beta-1)}{2}\Bigg)^{2}}\Bigg\{-8e^{2\mathcal{Z}}\upsilon^{2}\ln\Big(\frac{(\beta-2)e^{\mathcal{Z}}}{\beta-1}\Big)^{2}\times\nonumber\\&\!&\!
\times(\beta-2)^{2}\big(\beta^{4}-2\beta^{3}+\beta^{2}-\frac{3}{2}\big)-16\Bigg(\bigg((\lambda+\frac{3\upsilon}{2})\beta^{4}-(2\lambda+3\upsilon)\beta^{3}+(\lambda+\frac{3\upsilon}{2})\beta^{2}+-\frac{3}{4}(2\lambda+\upsilon)\bigg)(\beta-2)e^{2\mathcal{Z}}+\nonumber\\&\!&\!
+\frac{\Big(\beta^{3}-2\beta^{2}+\beta-\frac{3}{2}\Big)e^{\mathcal{Z}}(\beta-1)}{2}\Bigg)(\beta-2)\upsilon\ln\Big(\frac{(\beta-2)e^{\mathcal{Z}}}{\beta-1}\Big)-8\bigg((\lambda+\frac{3\upsilon}{2})^{2}\beta^{4}-(2\lambda^{2}+6\lambda\upsilon+\frac{13}{2}\upsilon^{2})\beta^{3}+(\lambda^{2}+\nonumber\\&\!&\!
+3\lambda\upsilon+\frac{25}{4}\upsilon^{2})\beta^{2}-2\beta\upsilon^{2}-\frac{3(\lambda+\frac{\upsilon}{2})^{2}}{2}\bigg)(\beta-2)^{2}e^{2\mathcal{Z}}-8\bigg((\beta-2)\Big((\lambda+\frac{5\upsilon}{2})\beta^{3}-(2\lambda+5\upsilon)\beta^{2}+(\lambda+\frac{5\upsilon}{2})\beta-\nonumber\\&\!&\!
-\frac{3}{4}(2\lambda+\upsilon)\Big)e^{\mathcal{Z}}-\frac{3}{8}(\beta-1)\bigg)(\beta-1)\Bigg\}.
\label{a8}    
\end{eqnarray}
The tensor-to-scalar ratio of the Logarithmic corrected model:
\begin{equation}
r=\frac{16\beta^{2}(\beta-2)^{2}\bigg(\upsilon\ln\Big(\frac{(\beta-2)e^{Z}}{\beta-1}\Big)+\lambda+\frac{3\upsilon}{2}\bigg)^{2}e^{2\mathcal{Z}}(\beta-1)^{2}}{3\bigg(e^{\mathcal{Z}}\upsilon(\beta-2)\ln\Big(\frac{(\beta-2)e^{\mathcal{Z}}}{\beta-1}\Big)+(\lambda+\frac{\upsilon}{2})(\beta-2)e^{\mathcal{Z}}+\frac{\beta-1}{2}\bigg)^{2}}.
\label{a9}    
\end{equation}
The running spectral index of the Logarithmic corrected model:
\begin{eqnarray}
&\!&\!\alpha_{s}=-\frac{1}{9\Bigg(e^{\mathcal{Z}}\upsilon(\beta-2)\ln\Big({\frac{(\beta-2)e^{\mathcal{Z}}}{\beta-1}}\Big)+(\lambda+\frac{\upsilon}{2})(\beta-2)e^{\mathcal{Z}}+\frac{(\beta-1)}{2}\Bigg)^{4}}\Bigg\{4\beta^{2}\Bigg(e^{\mathcal{Z}}\upsilon^{2}(\beta-2)(\beta-1)^{2}\ln\Big(\frac{(\beta-2)e^{\mathcal{Z}}}{(\beta-1)}\Big)^{2}+\nonumber\\&\!&\!
+2\bigg(-\beta\upsilon^{2}(\beta-2)^{2}e^{2\mathcal{Z}}+\Big(\frac{1}{4}+(\lambda+2\upsilon)(\beta-2)e^{Z}\Big)(\beta-1)^{2}\bigg)\upsilon\ln\Big(\frac{(\beta-2)e^{\mathcal{Z}}}{\beta-1}\Big)-2\Big((\frac{3\beta}{2}-2)n+\beta\lambda\Big)\upsilon^{2}(\beta-2)^{2}e^{2\mathcal{Z}}+\nonumber\\&\!&\!
+(\beta-1)\bigg(\Big(\big(\frac{15\beta}{4}-\frac{31}{4}\big)\upsilon^{2}+4\lambda\upsilon(\beta-1)+\lambda^{2}(\beta-1)\Big)(\beta-2)e^{\mathcal{Z}}+\frac{\big(\lambda+\frac{7\upsilon}{2}\big)(\beta-1)}{2}\bigg)\Bigg)(\beta-2)^{2}\bigg(\upsilon\ln{\Big(\frac{(\beta-2)e^{\mathcal{Z}}}{\beta-1}\Big)}+\nonumber\\&\!&\!
+\lambda+\frac{3\upsilon}{2}\bigg)e^{2\mathcal{Z}}(\beta-1)^{4}\Bigg\}
\label{a10}
\end{eqnarray}
where
\begin{equation}
\mathcal{Z}=\frac{2\upsilon\exp{\Big({\frac{8(1-\beta)^{2}\beta\upsilon N}{6e^{\frac{2\lambda+3\upsilon}{2\upsilon}}}}\Big)+3\upsilon+2\lambda}}{2\upsilon}.
\label{a11} 
\end{equation}
\end{document}